\documentclass[preprint2]{aastex}
\newcommand{\beq}{\begin{equation}}
\newcommand{\eeq}{\end{equation}}
\newcommand{\beqa}{\begin{eqnarray}}
\newcommand{\eeqa}{\end{eqnarray}}

\begin{document}

\title{The Cosmic Energy Inventory}

\author{Masataka Fukugita}

\affil{Institute for Advanced Study, Princeton, NJ 08540 USA}
\and\affil{Institute for Cosmic Ray Research, University of Tokyo,\\
            Kashiwa 277-8582, Japan}
\author  {P. J. E. Peebles}
\affil{Joseph Henry Laboratories,
Princeton University,\\
Princeton, NJ 08544 USA}

\begin{abstract}
We present an inventory of the cosmic mean densities of energy
associated with all the known states of matter and radiation at the
present epoch. The observational and theoretical bases for the
inventory have become rich enough to allow estimates with observational
support for the densities of energy in some 40 forms. The result is a
global portrait of the effects of the physical processes of cosmic
evolution.
\end{abstract}

\section{Introduction}
There is now a substantial observational basis for estimates of the
cosmic mean densities of all the known and more
significant forms of matter and energy in the present-day universe. The
compilation of the energy
inventory offers an overview of the integrated effects of the
energy transfers involved in all the physical processes of cosmic
evolution operating on scales ranging from the Hubble length to black
holes and atomic nuclei. The compilation also offers a way to assess
how well we understand the physics of cosmic evolution, by the degree
of consistency among related entries. Very significant
observational advances, particularly from large-scale surveys including
the {\it Two Degree Field Galaxy Redshift Survey}
(2dF: Colless et al. 2001), the {\it Sloan Digital Sky Survey}
(SDSS: York et al. 2000; Abazajian et al. 2003),
the {\it Two Micron All-Sky Survey} (2MASS: Huchra et al. 2003),
the {HI Parkes All Sky Survey} (HIPASS: Zwaan et al. 2003), and
the {\it Wilkinson Microwave Anisotropy Probe}
(WMAP: Bennett et al. 2003a), make it timely to compile what is known
about the entire energy inventory.

We present here our choices for the categories and estimates of the
entries in the inventory. Many of the arguments in this exercise are
updated versions of what is in the literature. Some arguments are new,
as is the adoption of a single universal unit --- the density parameter
--- that makes comparisons across a broad variety of forms of energy
immediate, but the central new development is that the considerable
range of consistency checks demonstrates that many of the entries in
the inventory are meaningful and believable.

People have been making inventories for a long time. The medieval
Domesday Book (1086-7) gave King William a picture of the wealth and 
organization of the kingdom (and it gives us fascinating insight into a society). The present-day cosmic energy inventory 
similarly gives us a picture of the amount and organization of the material contents of the universe. It also offers us a way to assess the 
reliability of our picture, through checks of consistency. A early 
example of the latter point is de~Sitter's (1917) discussion of 
Einstein's (1917) static homogeneous universe. Under the assumption 
that observations can reach a fair sample of the universe, one can seek 
constraints on the cosmic mean mass density and space curvature and 
test the predicted relation between the two. Hubble's (1929) 
redshift-distance relation led to a revision of the predicted relation between the mass density and space curvature, and his use of redshifts
to convert galaxy counts into number densities greatly improved
the estimate of the mean mass density (Hubble 1934).\footnote{Hubble's (1936) estimate based on galaxy masses derived from the velocity dispersion in the Virgo Cluster, which takes account of what is now termed nonbaryonic dark matter, translates to density parameter $\Omega _m\sim 0.1$, impressively close to the modern value.}
One may also consider the relations among the mean luminosity density
of the galaxies, the production
of the heavy elements, and the surface brightness of the night sky
(Partridge \& Peebles 1967; Peebles \& Partridge 1967); the relation
between galaxy colors, the initial mass function, and the star
formation history (Searle, Sargent, \& Bagnuolo 1973); the relation
between the last two sets of considerations (Tinsley 1973);  the
relation between the light element abundances and the baryon mass
density (Gott et al. 1974);
and the relation between the luminosity density of the quasars
and the mean mass density in quasars and their remnants (So\l tan
1982). Basu and Lynden-Bell (1990) show how one can analyze what is
learned from this rich set of considerations in terms of an entropy
inventory. We have chosen instead to base this discussion on an energy
inventory.

Our inventory includes the mass densities in the various states of
baryons. This is an updated version of the baryon budget of Fukugita,
Hogan \&\ Peebles (1998; FHP). Most entries in this part of the
inventory have not changed much in the past half decade, while
substantial advances in the observational constraints have considerably
reduced the uncertainties.  It appears that most of the baryonic
components are observationally well constrained, apart from the largest
entry, for warm plasma, which still is driven by the need to balance the
budget rather than more directly by the observations.

The largest entries, for dark matter and the cosmological constant, or
dark energy, are well constrained within a cosmological theory that is
reasonably well tested, but the physical natures of these entries
remain quite hypothetical. We understand the physical natures of
magnetic fields and cosmic rays, but the theories of the evolution of
these components, and the estimates of their contributions to the
present energy inventory, are quite uncertain. The situation for most of
the other entries tends to be between these extremes: the physical
natures of the entries are adequately characterized, for the most part,
and our estimates of their  energy densities, while generally not very
precise, seem to be meaningfully constrained by the observations.

Several cautionary remarks are in order. First, some types of energy
are not readily expressed as sums of simple components; we must adopt
conventions. Second, there is no arrangement of categories that offers
a uniquely natural place for each component; we must again adopt
conventions. Perhaps further advances in the understanding of cosmic
evolution will lead to a more logically ordered inventory.
Third, it is arguably artificial to represent binding energies as very
small negative density parameters. The advantage is that
it simplifies comparisons across the entire inventory.
Fourth, the amount of space we devote is in accord with the importance 
of the issues of physics and astronomy, not with the value of the 
energy. For instance, the major
components in the inventory are dark energy and dark matter, but they 
are physically simple in that
their work is only gravitational, so our discussion is rather short.
Dissipative energies are small in absolute values, but their
physical significance is large, and our discussions are on occasion 
rather lengthy. Finally, it is a task for future work to make some of  
our estimates
more accurate by using data and computations that exist but are
difficult to assemble. We mention the main examples in  \S 3.

The inventory, which is presented in Table~1, is arranged by categories
and components within categories. The explanations of conventions and
sources for each entry are presented in \S 2, in subsections with
numbers keyed to the category numbers in the first column of the table.
Our discussion of checks of the entries is not so simply ordered,
because the checks depend on relations among considerations of
entries scattered through the table. A guide to  the considerable
variety of checks detailed in \S 2 is presented in \S 3.

\begin{deluxetable}{lllllr}
\tabletypesize{\scriptsize}
\tablecaption{The Cosmic Energy Inventory \label{tbl-1}}
\tablewidth{0pt}
\tablehead{ \colhead{}  & \colhead{}  &  \colhead{} & \colhead{} &
\colhead{Components\tablenotemark{a} }  &
\colhead{Totals\tablenotemark{a}}}
\startdata
1 & \multicolumn{3}{l} {dark sector} & &  $0.954\pm 0.003$ \\
1.1  & & dark energy   & & $0.72\pm 0.03$  \\
1.2 & & dark matter   & & $0.23 \pm 0.03$ \\
1.3 & &\multicolumn{2}{l}{primeval gravitational waves} &   $\la
10^{-10}$ \\
\noalign{\vspace{3pt}}\tableline\noalign{\vspace{3pt}}
2 & \multicolumn{4}{l}{primeval thermal remnants} & $0.0010\pm 0.0005$
\\
2.1 &&\multicolumn{2}{l}{electromagnetic radiation} & $10^{-4.3\pm 
0.0}$ \\
2.2 & &\multicolumn{2}{l}{neutrinos} & $10^{-2.9\pm0.1}$ \\
2.3 &&\multicolumn{2}{l}{prestellar nuclear binding energy} &
$-10^{-4.1\pm 0.0}$ \\
\noalign{\vspace{3pt}}\tableline\noalign{\vspace{3pt}}
3 & \multicolumn{3}{l} {baryon rest mass} & & $0.045\pm 0.003$ \\
3.1 & &\multicolumn{2}{l}{warm intergalactic plasma}
&\multicolumn{2}{l}{$0.040\pm 0.003$}\\
3.1a & &\multicolumn{1}{l}{\qquad virialized regions of
galaxies\hspace{-0.5in}}
&\multicolumn{1}{l}{\qquad\quad $0.024\pm 0.005$}\\
3.1b & &\multicolumn{1}{l}{\qquad intergalactic}
&\multicolumn{1}{l}{\qquad\quad $0.016\pm 0.005$}\\
3.2 & &\multicolumn{2}{l}{intracluster
plasma }&\multicolumn{2}{l}{$0.0018\pm 0.0007$}\\
3.3 & & main sequence stars\qquad\qquad  &  spheroids and bulges &
\multicolumn{2}{l}{$0.0015\pm 0.0004$}\\
3.4 &&&disks and irregulars & \multicolumn{2}{l}{$0.00055\pm 0.00014$}\\
3.5 & &white dwarfs  &&\multicolumn{2}{l}{$0.00036\pm 0.00008$}\\
3.6 & &neutron stars  &&\multicolumn{2}{l}{$0.00005\pm 0.00002$}\\
3.7 & &black holes  &&\multicolumn{2}{l}{$0.00007\pm 0.00002$}\\
3.8 & &substellar objects&&\multicolumn{2}{l}{$0.00014\pm 0.00007$}\\
3.9 & &HI + HeI  & & \multicolumn{2}{l}{$0.00062\pm 0.00010$}\\
3.10 & &molecular gas & &\multicolumn{2}{l}{$0.00016\pm 0.00006$}\\
3.11 & & \multicolumn{2}{l}{planets}
&\multicolumn{1}{l}{$10^{-6}$} \\
3.12 & & \multicolumn{2}{l}{condensed matter}
&\multicolumn{1}{l}{$10^{-5.6\pm 0.3}$} \\
3.13 & & \multicolumn{2}{l}{sequestered in massive black holes}
&\multicolumn{1}{l}{$10^{-5.4}(1 + \epsilon _n)$} \\
\noalign{\vspace{3pt}}\tableline\noalign{\vspace{3pt}}
4 & \multicolumn{4}{l} {primeval gravitational binding energy} &
$-10^{-6.1\pm 0.1}$ \\
4.1 & & \multicolumn{2}{l}{virialized halos of galaxies}  &$-10^{-7.2}$
\\
4.2 & & clusters &  &$-10^{-6.9}$\\
4.3 & &  \multicolumn{2}{l}{large-scale structure}  &$-10^{-6.2}$\\
\noalign{\vspace{3pt}}\tableline\noalign{\vspace{3pt}}
5 & \multicolumn{3}{l} {binding energy from dissipative gravitational
settling} & & $ -10^{-4.9} $ \\
5.1 & &\multicolumn{2}{l}{baryon-dominated parts of
galaxies} &$-10^{-8.8\pm 0.3}$\\
5.2 & & \multicolumn{2}{l}{main sequence stars and substellar objects}
&$-10^{-8.1}$\\
5.3 && white dwarfs &&$-10^{-7.4}$\\
5.4 &&neutron stars    &&$-10^{-5.2}$\\
5.5 && \multicolumn{2}{l}{stellar mass black holes} &
$-10^{-4.2}\epsilon_s$  \\
5.6 &&  galactic nuclei  & early type &  $-10^{-5.6}\epsilon _n$   \\
5.7 && & late type & $-10^{-5.8} \epsilon _n$   \\
\noalign{\vspace{3pt}}\tableline\noalign{\vspace{3pt}}
6.  & \multicolumn{4}{l} {poststellar nuclear binding energy} &
$-10^{-5.2}$ \\
6.1 & & \multicolumn{2}{l}{main sequence stars and substellar objects} &
$-10^{-5.8}$\\
6.2 & & \multicolumn{2}{l}{diffuse material in galaxies} &$-10^{-6.5}$\\
6.3 & & \multicolumn{2}{l}{white dwarfs} &$-10^{-5.6}$\\
6.4 & & clusters &&$-10^{-6.5}$\\
6.5 && intergalactic &&$-10^{-6.2\pm0.5}$\\
\noalign{\vspace{3pt}}\tableline\noalign{\vspace{3pt}}
7 & \multicolumn{4}{l} {poststellar radiation} &  $10^{-5.7\pm0.1}$ \\
7.1 & & \multicolumn{2}{l} {resolved radio-microwave} &$10^{-10.3\pm 
0.3}$\\
7.2 & & far infrared &&$10^{-6.1}$\\
7.3 &  & optical &&$10^{-5.8\pm0.2}$\\
7.4&& X-$\gamma$ ray &&$10^{-7.9\pm 0.2}$\\
7.5&& gravitational radiation & stellar mass binaries &$ 10^{-9\pm 1 }$\\
7.6&&  & massive black holes &$ 10^{-7.5\pm 0.5}$\\
\noalign{\vspace{3pt}}\tableline\noalign{\vspace{3pt}}
8 & \multicolumn{4}{l} {stellar neutrinos}  &  $10^{-5.5}$\\
8.1 & & \multicolumn{2}{l}{nuclear burning} & $10^{-6.8}$ \\
8.2 & &  \multicolumn{2}{l}{white dwarf formation} & $10^{-7.7}$ \\
8.3 && \multicolumn{2}{l}{core collapse} & $10^{-5.5}$ \\
\noalign{\vspace{3pt}}\tableline\noalign{\vspace{3pt}}
9 &\multicolumn{4}{l}{cosmic rays and magnetic fields} &
$10^{-8.3^{+0.6}_{-0.3}}$\\
\noalign{\vspace{3pt}}\tableline\noalign{\vspace{3pt}}
10 &\multicolumn{4}{l}{kinetic energy in the intergalactic medium} &
$10^{-8.0\pm 0.3}$\\
\enddata
\tablenotetext{a}{Based on Hubble parameter $h=0.7$.}
\end{deluxetable}

\section{The Energy Inventory}

The inventory in Table~1 assumes the now standard relativistic
Friedmann-Lema\^\i tre $\Lambda$CDM cosmology, in which space sections
at fixed world time have negligibly small mean curvature,\footnote{
The most compelling evidence is from the position of the first peak in
the anisotropy power spectrum of the of cosmic microwave background 
temperature field.
The  WMAP measurements (Bennett et al. 2003a) indicate
$\Omega_m+\Omega_\lambda=1.02 \pm 0.02$.
In this paper we simplify the discussion by assuming strictly flat 
space sections.
} Einstein's
cosmological term, $\Lambda$, is independent of time and position, the
dark matter is an initially cold noninteracting gas, and primeval
departures from homogeneity are adiabatic, Gaussian, and
scale-invariant. Physics in the dark sector is not well constrained:
$\Lambda$ might be replaced with a dynamical component,\footnote{
The current limit on the index of the equation
of state for the dark energy is $w=p/\rho<-0.78$ at 95\% (Bennett 
2003a); the bound
$w=-1.02^{+0.13}_{-0.19}$ is obtained from the Type Ia supernova
Hubble diagram under the assumption of flat space curvature
(Riess et al. 2004)} as in the
models for dark energy now under discussion, the physics of the dark
matter may prove to be more complicated than that of a free
collisionless gas, and the initial conditions may not be adequately
approximated by the present standard cosmology. If such complications
were present we expect their effects on entries that are sensitive to
the cosmological model would be slight, however, because the
cosmological tests now offer close to compelling evidence that the
$\Lambda$CDM model is a useful approximation to reality (Bennett et al.
2003a; Spergel et al. 2003; Tegmark et al. 2004a; and references
therein).

To help simplify the discussion we adopt a nominal distance scale,
corresponding to Hubble's constant\footnote{We also write
$H_o=100h\hbox{ km s}^{-1}\hbox{ Mpc}^{-1}$, where convenient, but all
entries in the inventory in Table~1 assume $h=0.7$.}
\beq
H_o = 70 \hbox{ km s}^{-1}\hbox{ Mpc}^{-1}.\label{eq:hnot}
\eeq
The energy density, $\bar\rho _i$, in the form of component $i$ is
expressed as a density parameter,
\beq
\Omega _i = {8\pi G\bar\rho _i\over 3H_o^2}.
\eeq
Since $H_o$ is at best measured to ten percent accuracy (Freedman et
al. 2001), an improved distance scale could produce noticeable
revisions to the inventory.

The second column from the right in Table~1 lists the density
parameters in the components, and the last column presents the total
for each category. Both columns sum to unity. The statement of errors 
requires some explanation. Most of the errors in categories 1 through 3 
are rather well documented. The uncertainties in entries 2.1 and 2.3 
are too small to be relevant for the purpose of this inventory. 
Elsewhere a numerical value stated to one significant figure after the 
decimal might be supposed to be reliable to roughly $\pm 0.1$~dex, or 
about 30\%. Where no digit is presented after the decimal point we hope 
our estimate might within a factor of ten of the
real value.

\subsection{The Dark Sector}
\label{sec:darksector}

The components in category 1 interact with the contents of the visible
sector only by gravity, as far as is now known. This makes it difficult
to check whether the dark energy --- or Einstein's cosmological
constant, $\Lambda$ --- and the dark matter really have the simple
properties
assumed in the $\Lambda$CDM cosmology. Future versions of the
inventory might contain separate entries for the potential, kinetic  and
gradient contributions to the dark energy density, or a potential
energy component in the dark matter.

There is abundant evidence that the total mass density --- excluding
dark energy --- is well below the Einstein-de Sitter value. That means,
among other things, that the consistency of cross-checks from the many
ways to estimate the mass density provides close to compelling evidence
that the gravitational interaction of matter at distances up to the
large-scale flows is well approximated by the inverse square law, and
that starlight is a good tracer of the mass distribution on scales $\ga
100$~kpc.\footnote{For early discussions see Peebles 1986 and Bahcall, 
Lubin \&\ Dorman (1995). Recent reviews of the situation are in 
Fukugita (2001), Peebles \& Ratra (2003), and Bennett et al. (2003a). 
The present observational situation is reviewed in footnote 6.}

An example that illustrates the situation, and will be used later, is
the estimate from weak lensing of the mean galaxy surface mass density
contrast,
\beq
\Sigma (<y) - \Sigma (y) = A(hy/1\hbox{
Mpc})^{-\alpha},\label{eq:delatsigma}
\eeq
where $\Sigma (y)$ is the ensemble average surface mass density at
projected distance $y$ from a galaxy and $\Sigma (<y)$ is the mean
surface density within distance $y$. The measurements by McKay et al.
(2001) yield $A = 2.5_{-0.8}^{+0.7}\, h\, m_\odot \hbox{ pc}^{-2}$ and
$\alpha = -0.8\pm 0.2$, and they indicate that the power law is a good
approximation to the measurements in the range of projected radii
\beq
70\hbox{ kpc}\la y \la 1\hbox{ Mpc}.\label{eq:rangeofscales}
\eeq
If the galaxy autocorrelation function,
\beq
\xi (r) = \left(r_o\over r\right) ^\gamma, \ \gamma =1.77, \
r_o=5h^{-1}\hbox{ Mpc},\label{eq:galaxyxi}
\eeq
is a good approximation to the galaxy-mass cross correlation function
on the range of scales in equation~(\ref{eq:rangeofscales}), then the
mean surface density is
\beqa
\Sigma (y) &=& \rho _m r_o^\gamma y^{1 -\gamma }H_\gamma,\nonumber\\
H_\gamma &=&  {\Gamma (1/2)\Gamma ((\gamma - 1)/2)\over\Gamma (\gamma
/2)} .
\eeqa
This agrees with equation~(\ref{eq:delatsigma}) if the density
parameter belonging to the mean density $\rho _m$ of the mass
that clusters with the galaxies is
\beq
\Omega _m({\rm weak\ lensing}) =
0.20^{+0.06}_{-0.05}.\label{eq:weaklensing}
\eeq
This  is in the range of estimates of the value of this parameter
now under discussion, consistent with the assumption that galaxies are 
useful tracers of mass.

Our adopted value for the total mass density in
nonrelativistic matter, dark plus baryonic, is
\beq
\Omega _m = \Omega _{\rm DM} + \Omega _{\rm b} +
\Omega _\nu = 0.28\pm 0.03.
\label{eq:omegam}
\eeq
This is in the range of most current 
estimates.\footnote{Spergel et al. derived $\Omega_m h^2=0.13-0.14$
from WMAP either with or without using the constraint 
from the power spectrum of 
the 2dF galaxy distribution. This gives $\Omega_m=0.265-0.286$
at our fiducial $h=0.7$. Tegmark et al. (2004a) obtained the central 
value
$\Omega_m h^2=0.14$ with WMAP data alone, and 0.145 with the constraint 
  from the SDSS galaxy clustering. For $h=0.7$ these values are 
respectively
$\Omega_m=0.29$ and 0.30. The estimate from the 2dFGRS power spectrum 
(W. Percival and the 2DFGRS team, 2004, private communication) is 
$\Omega_m h=0.164\pm0.016$, or $\Omega _m=0.23\pm 0.02$ at  our 
distance scale. This is 1.4 standard deviations below equation 
(\ref{eq:omegam}). The lower estimate of $\Omega _m$ is due to the somewhat larger value of the small-scale power spectrum compared to SDSS. Note that departures from the standard assumptions, 
including flat space curvature, scale-invariant initial conditions, and 
neglible tensor perturbations, could lead to
changes beyond the quoted errors.
We also refer to estimates from the Type Ia sypernova
redshift-magnitude relation, $\Omega_m=0.29^{+0.05}_{-0.03}$ (Riess et 
al. 2004), and from dynamics, including $\Omega _m=0.17\pm 0.05$ from 
the cluster mass function as a function of redshift (Bahcall \& Bode 
2003) and $\Omega _m=0.30\pm 0.08$ from the redshift space two-point correlation function (Hawkins et al. 2003), under the assumption that the bias parameter is $b=1$ (Verde et al. 2002).} The 
measurement may not be tightly constrained, however, 
and there is the possibility of 
adjustment of this important parameter beyond the error flag in
equation~(\ref{eq:omegam}).

We make use of the fact that equation (\ref{eq:delatsigma})
is close to the limiting isothermal sphere mass distribution,
\beq
\rho(r)=\sigma^2/2\pi Gr^2
\label{eq:isothermal}
\eeq
If we connect this form to the power law
in equation~(\ref{eq:galaxyxi})
at the nominal virial radius $r_v$ defined by
$\rho(<r_v)/\rho_c=200$, we obtain
\beq
r_v=220h^{-1}{\rm kpc},~~~~\sigma=160 \hbox{ km}~s^{-1}.
\label{eq:virial}
\eeq
This measure of the characteristic one-dimensional velocity dispersion,
$\sigma$, in luminous galaxies agrees with the mean in quadrature,
\beq
\langle\sigma^2\rangle^{1/2}\simeq 160\hbox{ km s}^{-1},
\eeq
weighted by the FHP morphology fractions, of the dispersions
$\sigma=225~$km~s$^{-1}$ for elliptical galaxies,
$\sigma=206~$km~s$^{-1}$ for S0 galaxies
(de Vaucouleurs \& Olson 1982) and $\sigma=136~$km~s$^{-1}$
for spiral galaxies (Sakai et al. 2000), all at luminosity $L_B=L_B^*$.
The isothermal sphere model defines a characteristic mass $M_v$ within
$r_v$. The ratio of $M_v$ to  the characteristic galaxy luminosity
(eq.~[(\ref{eq:charlum}] below) is
\beq
M_v/L_r = 180h,\quad M_v/L_B=250h,
\label{eq:masstolight}
\eeq
in Solar units.
This is consistent with the estimate of $M/L$ within 220$h^{-1}$
kpc from weak lensing shear (McKay et al. 2001).\footnote{It is worth 
noting that equation~(\ref{eq:masstolight}) is not far from   Zwicky's 
(1933) dynamical estimate for the Coma Cluster, $M/L\sim 100$ at our 
distance scale.}

The product of  $M_v$ with the effective number density
of luminous ($L^*$) galaxies, $n_g= 0.017h^3$ Mpc$^{-3}$
(eq~[\ref{eq:galdensity}]),
gives an estimate that the mean mass fraction within the virial
radii of the large galaxies,
\beq
{\rho(<r_v) \over \rho_m}=0.6.
\label{eq:galaxydarkmass}
\eeq
That is, we estimate that about 60\%  of the dark matter is gathered
within the virialized parts of normal galaxies.

We consider now how the estimate of $\Omega _m$ in 
equation~(\ref{eq:omegam}) compares to the estimate from the 
mass-to-light ratio and the integrated mean luminosity density. The 
estimates from the SDSS broad-band galaxy luminosity functions are 
(Yasuda et al. in preparation; see also
Blanton et al. 2003)
\beqa
{\cal L}_B&=&(1.9\pm 0.2)\times10^8hL_\odot\hbox{Mpc}^{-3}, \nonumber\\
{\cal L}_r&=&(2.3\pm 0.2)\times10^8hL_\odot\hbox{Mpc}^{3},
\label{eq:luminositydensity}\\
{\cal L}_z&=&(3.6\pm 0.4)\times10^8hL_\odot\hbox{Mpc}^{-3}.\nonumber
\eeqa
These densities are the values at $z\approx 0.05$. The value of
${\cal L}_B$ is inferred from the densities in other color
bands.\footnote{The ratios of luminosity densities, 1 : 1.20 : 1.87, 
are approximately what is expected  from the average colors, $\langle 
B-r \rangle=1.0$ and $\langle r-z \rangle=0.6$ [$(B-r)_\odot=0.82$, 
$(r-z)_\odot=0.12$].} The luminosity density in the $z$ band is quoted 
for the later use. The product of the luminosity density with $M/L$ in equation (\ref{eq:masstolight}) yields the density parameter in matter within the virial radii of galaxies, $\Omega_{m,v}=0.18$ and 0.16
for the $B$ and $r$ bands, respectively. On dividing by  equation
(\ref{eq:galaxydarkmass}) we
arrive at $\Omega_m=0.31$ and $0.27$,  consistent with equation
(\ref{eq:omegam}).

Note that $M/L$ in equation (\ref{eq:masstolight}) is about half
the value estimated for rich clusters,  $(M/L)_{\rm cluster}=450 \pm 
100$ (FHP). In rich clusters all mass in the outskirts of galaxies, 
beyond their
virial radii, is integrated in the mass estimate, so that it is
reasonable to suppose that it
gives a larger value by the inverse factor in equation 
(\ref{eq:galaxydarkmass}).
When the cluster value for $M/L$ is multiplied by the $B$ band 
luminosity
density we get $\Omega_m=0.32$, consistent with the other estimates.

Entry 1.3 assumes inflation has produced gravitational waves with a
scale-invariant spectrum, meaning the strain $\delta$ appearing at the
Hubble length is independent of epoch. The density parameter associated
with gravitational waves with wavelengths on the order of the Hubble
length is $\Omega_g\sim\delta ^2$, and the absence of an appreciable
effect of gravitational waves on the anisotropy of the 3~K thermal
cosmic background radiation indicates $\delta\la 10^{-5}$.
The gravitational waves produced by cosmic phase transitions, if
detected
or convincingly  predicted, might be entered in this category.
Gravitational waves from the relativistic collapse of stars and
galactic nuclei  are included in category~7.

The other entries in the first category in Table~1 are computed from
equation~(\ref{eq:omegam}) and our estimates of the other significant
contributions to the total mass density, under the assumption that the
density parameters sum to unity, that is, space curvature is
neglected.

\subsection{Thermal Remnants}

\subsubsection{Cosmic Background Radiation}

Entry 2.1 is based on the COBE measurement of the temperature of the
thermal cosmic electromagnetic background radiation (the CMBR),
$T_o=2.725$~K (Mather et al. 1999).
The COBE and UBC measurements (Mather et al. 1990; Gush, Halpern, \&
Wishnow 1990) show that the spectrum is very close to thermal. It
has been slightly disturbed by the observed
interaction with the hot plasma in
clusters of galaxies (LaRoque et al. 2003 and references therein). The
limit on the resulting fractional increase in the radiation energy
density is (Fixsen et al. 1996)
\beq
\delta u/ u =4y < 6\times 10^{-5}.\label{eq:y-parameter}
\eeq
This means that the background radiation energy density has been
perturbed by the amount $\Delta\Omega < 10^{-8.5}$. Improvements of this
number are under discussion (e.g. Zhang, Pen \&\ Trac 2004), and might
be entered in a future version of the inventory.

The thermal background radiation has been
perturbed also by the dissipation of the primeval fluctuations in the
distributions of baryons and radiation on scales smaller than the
Hubble length at the epoch of decoupling of  baryonic matter and
radiation. If the initial  mass fluctuations are adiabatic and
scale-invariant the fractional perturbation to the radiation energy per
logarithmic increment of the comoving length scale is $\delta
u/u\sim\delta _h^2$, where $\delta _h\sim 10^{-5}$ is the density
contrast appearing at the Hubble length. This is small compared to the
subsequent perturbation by hot plasma (eq~[\ref{eq:y-parameter}]).

Entry 2.2 uses the standard estimates of the relict thermal neutrino
temperature, $T_\nu = (4/11)^{1/3}T_o$, and the number density  per
family, $n_\nu = 112$~cm$^{-3}$. We adopt the neutrino mass differences
   from oscillation experiments (Fukuda et al. 1998; Kameda et al. 2001;
Eguchi, et al. 2003; Bahcall \&\ Pe\~na-Garay 2003),
\beqa
&& m_{\nu_3 }^2 - m_{\nu _2 }^2 = 0.002\pm 0.001 \hbox{
eV}^2,\nonumber\\
&&m_{\nu _2 }^2 - m_{\nu _1}^2 = 6.9\times 10^{-5}\hbox{ eV}^2,
\eeqa
where the neutrino mass eigenstates are ordered as
$m_{\nu _1}<m_{\nu _2}<m_{\nu _3}$.
Entry 2.2, the density parameter $\Omega _\nu$ in primeval neutrinos,
assumes that $m_{\nu_e}$ may be neglected. The upper limit from WMAP
and SDSS is $\Omega _\nu < 0.04$ (Tegmark et al. 2004a). At this limit
the
three families would have almost equal masses, $m_\nu =0.6$~eV.
This may not be very likely, but one certainly must bear in mind the
possibility that our entry is a considerable underestimate.

\subsubsection{Primordial Nucleosynthesis}

Light elements are produced as the universe expanded and cooled
through $kT\sim 0.1$~MeV, in amounts that depend
on the baryon abundance. The general agreement of the baryon
abundance inferred in this way with
that derived from the CMBR
temperature anisotropy gives confidence that
the total amount of baryons --- excluding what might have been trapped
in the dark matter prior to light element nucleosynthesis --- is
securely constrained.

Estimates of the baryon density parameter from the WMAP and SDSS data
(Spergel et al. 2003;
Tegmark et al. 2004a), and from the deuterium (Kirkman et
al. 2003) and helium abundance measurements (Izotov \& Thuan 2004)
are, respectively,
$\Omega _{\rm b}h^2 =0.023\pm 0.001$, $0.0214\pm 0.0020$, and
$0.013^{+.002}_{-0.001}$, where the last number is the all-sample
average for helium from Izotov \& Thuan. We adopt
\beq
\Omega _{\rm b}h^2 =0.0225\pm 0.0015,\label{eq:baryonomega}
\eeq
close to the mean of the first two. Since the relation between the
helium abundance and the baryon density parameter has a very shallow
slope, an accurate abundance estimate (say, with $<1$\% error) is
needed for a strong constraint on $\Omega _{\rm b}h^2$. We consider
that the current estimates may still suffer from
systematic errors which are not included in the error
estimates in the literature.\footnote{We note, as an indication of the
difficulty of these observations, that helium abundances
inferred from the triplet 4d-2p transition
($\lambda 4471$) are lower than what is indicated by the triplet 3d-2p
($\lambda 5876$) and singlet 3d-2p  ($\lambda 6678$)
transitions, by an amount that is significantly larger than
the quoted errors. Another uncertainty arises from
stellar absorption corrections, which are calculated only for
the $\lambda 4471$ line. The table given in Izotov and Thuan
suggests that a small change in absorption corrections for
the $\lambda 4471$ line induces a sizable change in the final
helium abundance estimate.  We must remember also that the value of 
$\Delta Y/\Delta Z$, which is needed to derive $Y_p$, is not
very well determined.}
Within the standard cosmology our adopted value in
equation~($\ref{eq:baryonomega}$) requires that the primeval helium
abundance is
\beq
Y_p=0.248\pm 0.001,\label{eq:primevalhelium}
\eeq
and the ratio of the total matter density to the baryon component is
\beq
     \Omega _m/\Omega_b=6.11\pm0.89 .\label{eq:omoverob}
\eeq

We need in later sections the stellar helium production rate
with respect to that of the heavy elements.
The all-sample analysis of Izotov \& Thuan (2004) gives
$\Delta Y/\Delta Z \simeq 2.8\pm0.5$. The value derived by
Peimbert, Peimbert \& Ruiz (2000) corresponds to $2.3\pm0.6$.
These value may be compared to estimates from
the perturbative effects on the effective temperature-luminosity
relation for the atmosphere of main sequence dwarfs,
$\Delta Y/\Delta Z \simeq 3\pm2$ (Pagel \& Portinari 1998),
and $2.1\pm0.4$ (Jimenez et al. 2003). From the initial elemental
abundance estimate in the standard solar
model of Bahcall, Pinsonneault \& Basu (2001; hereinafter BP2000) we
derive
$\Delta Y/\Delta Z \simeq 1.4$.
We adopt
\beq \Delta Y/\Delta Z \simeq 2\pm 1 .
\label{eq:dydz}
\eeq

Nuclear binding energy was released during nucleosynthesis. This
appears in entry 2.3 as a negative value, meaning the comoving
baryon mass density has been reduced and the energy density in
radiation and neutrinos increased. The effect on the radiation
background has long since been thermalized, of course, but the entry is
worth recording for comparison to the nuclear binding energy released
in stellar evolution. For the same reason, we compute the binding
energy relative to free protons and electrons. The convention is
artificial,
because light element formation at high redshifts was
dominated by radiative exchanges of neutrons, protons  and atomic
nuclei, and the abundance of the neutrons was determined by energy
exchanges with the cosmic neutrino background. It facilitates
comparison with category~6, however.
The nuclear binding energy in entry 2.3 is the product
\beq
-\Omega _{\rm NB, He} = 0.0071\, Y_p\,\Omega_{\rm b}=10^{-4.1}.
\eeq
This is larger in magnitude than the energy in the CMBR today.

\subsection{The Baryon Rest Mass Budget}

The entries in this category refer to the baryon rest mass: one must
add the negative binding energies to get the present mass density in
baryons. The binding energies are small and the distinction purely
formal at the accuracy we can hope for in cosmology, of course, with
the conceivable exception of the baryons sequestered in black holes.

We begin with the best-characterized components, the stars,
star remnants, and planets. We then consider the diffuse components, and
conclude this subsection with discussions of the baryons in
groups and the intergalactic medium and the lost baryons in black holes.

\subsubsection{Stars}
This is an update of the analysis in FHP.
Following the same methods, we estimate the baryon mass in stars from
the galaxy luminosity density and the stellar mass-to-light ratio,
$M_{\rm stars}/L$, along with a stellar initial mass function (the IMF)
that allows us to
estimate the mass fractions in various forms of stars and star remnants.

Kauffmann et al. (2003) present an extensive analysis of the stellar
mass-to-light ratio based on $ugriz$ photometry for
$10^5$ SDSS galaxies and a population
synthesis model (Bruzual \& Charlot 2003)
that is meant to take account of the stellar
metallicities and the star formation histories.
Their estimate of the stellar mass-to-light ratio
is $M_{\rm stars}/L_z\simeq 1.85$ for luminous galaxies, with magnitudes
$M_z<M_z^*-0.8$, and it decreases gradually to
$M_{\rm stars}/L_z\simeq 0.65$ for
fainter galaxies with $M_z\simeq M_z^*+3$ for a galaxy sample
at $z\approx 0.05$. Our estimate of the
resulting luminosity-function-weighted mean is
\beq
\langle M_{\rm stars}/L_z\rangle\simeq1.5\pm 0.3,
\label{eq:kauffmannmol}
\eeq
for the IMF Kauffmann et al. used. Equation~(\ref{eq:kauffmannmol})
represents the present mass in stars and stellar remnants, and it
excludes the mass shed by evolving stars and returned to diffuse
components.

The estimate of $M_{\rm stars}/L$ assumes a universal IMF,
which is not thought to seriously violate
the observational constraints. We note, however, that a possible
change of the IMF at very high redshift need not seriously affect our 
analysis because star formation a high redshift contributes little to 
the present total mass in stars. The IMF is particularly  uncertain at 
the subsolar masses that make little
contribution to the light but can make a considerable contribution to
the mass. We consider two continuous broken power law models, of the
form
\beq dN/dm \propto m^{-(x+ 1)},\eeq
where, in the first model,
\beqa
x &=&  -0.5, \quad  0.01 < m < 0.1 m_\odot,\nonumber\\
 ~ &=& 0.25\quad 0.1 < m < 1 m_\odot,\label{eq:imfreid}\\
 ~ &=& 1.35\quad 1 < m < 100 m_\odot,\nonumber
\eeqa
and, in the second model,
\beqa
x &=&  -0.7\quad  0.01 < m < 0.08 m_\odot,\nonumber\\
 ~ &=& 0.3\quad 0.08 < m < 0.5 m_\odot,\label{eq:imfkroupa}\\
 ~ &=&  1.3\quad 0.5 < m < 100 m_\odot .\nonumber
      \eeqa
The first line in the first model is from Burgasser et al. (2003),
the second line is from Reid, Gizis, \& Hawley (2002), and the third line,  for $m>1 m_\odot$, is the standard Salpeter (1955) IMF.
The second model is from Kroupa (2001). Yet another IMF, that of Chabrier (2003), is in between these two. For our model IMF we take the Salpeter index for $m\ge 1 m_\odot$. It is known that the Salpeter slope gives satisfactory UBV colors and H$\alpha$
equivalent widths for normal galaxies (Kennicutt 1983),
whereas an IMF with a steeper slope (e.g., Scalo 1986) is not favored in this regard.  The consensus seems to be that the power law index is smaller than the Salpeter value at sub-Solar masses. The two IMF given above still differ significantly at $m<m_\odot$, however,  For the subsolar IMF, we take the geometric
mean of the above two models after mass integration, and we take the
difference as an indication of the error ($\pm 18$\%).

\begin{deluxetable}{lcccc}
\tabletypesize{\scriptsize}
\tablecaption{Star mass fractions}
\tablewidth{0pt}
\tablehead{ \colhead{initial mass range}  & \colhead{fate}  &
\colhead{remnant\tablenotemark{a}} & \colhead{mass fraction} &
\colhead{mass consumed\tablenotemark{b}}}
\startdata
$0.01< m < 0.08$ & SS & \nodata &  0.052 & 0.052\\
$0.08 < m < 100$\tablenotemark{c} & MS & \nodata &  0.769 & 0.769\\
$1 < m < 8$ & WD & 0.62 & 0.135 & 0.463\\
$8 < m < 25$ & NS & 1.35 & 0.019 & 0.186\\
$25 < m < 100$ & BH & 7.5 & 0.025 & 0.146\\
\noalign{\vspace{3pt}}\tableline\noalign{\vspace{3pt}}
sum       &   &   &  1.0  &  1.616 \\
\enddata
\tablenotetext{a}{Mass in units of Solar masses}
\tablenotetext{b}{The gas consumed to make unit mass of stars
now present.}
\tablenotetext{c}{For the mass range $1 < m < 100 m_\odot$ those
stars burning today are counted.}
\end{deluxetable}

The stellar mass-to-light ratio in equation~(\ref{eq:kauffmannmol})
assumes the IMF in equation~(\ref{eq:imfkroupa}). With our adopted IMF
the stellar mass-to-light ratio is 1.18 times
the number in equation~(\ref{eq:kauffmannmol}).\footnote{The
IMF are normalised so that the mass integrals between 0.9 and
2.0$m_\odot$
are equal. The result is virtually identical with those with the
two IMFs normalised at 1$m_\odot$.} Thus we get our fiducial estimate,
\beq
\langle M_{\rm star}/L_z\rangle = 1.23\pm 0.33.\label{eq:ourimf}
\eeq
This translates to $M/L_B\approx 2.4$, or 0.7 times that
used in FHP, which employed the subsolar mass IMF of
Gould, Bahcall \& Flynn (1996). In the Salpeter IMF, with $x = 1.35$
cut off at $0.1m_\odot$, the mass-to-light ratio is 1.48 times our
adopted value. The Kennicutt (1983) IMF results in 0.81 times
equation~(\ref{eq:ourimf}).

The IMF at substellar masses,
$m<0.08m_\odot$, is more
uncertain, but recent observations of T dwarfs in the solar
neighbourhood indicate $x<0$
(e.g., Burgasser et al. 2003; 2004).
In the two IMF models quoted above the substellar mass is
6 to 9\% of the mass integral for
$0.08<m<1m_\odot$. We adopt 8\% and assign an error of 50\%.

We estimate the mass density locked in stars, including those
in dead stars, to be $\Omega _{\rm stars}=0.0024\pm0.0007$.
A comparable estimate is derived from $b_J,J,K_s$ multicolour
photometry of 2MASS combined with 2dF data by  Cole et al. (2001),
$\Omega _{\rm stars}= 0.0029 \pm 0.0004$ with the
IMF we have adopted.
For the energy inventory we
take the mean of our present number and that of Cole et al.:
\beq
\Omega _{\rm stars}=0.0027 \pm 0.0005.\label{eq:omegastars}
\eeq
This means that the stars contain $6.0\pm1.3$\% of the total baryons.
The FHP estimate is $\Omega_{\rm stars}=0.0019-0.0057$.
Equation~(\ref{eq:omegastars}) also is consistent within the
errors with the more recent estimates by Salucci \& Persic (1999),
Kochanek et al. (2001), and Glazebrook et al. (2003), and with Shull's
(2003) baryon inventory.

We attempt to partition the stars into their species.
Our estimates of the mass fractions in stars on the main sequence
(MS) and substellar objects (SS),
and the mass fractions in stellar remnants, including white
dwarfs (WD), neutron stars (NS) and stellar mass black holes (BH), are
shown in Table~2. Stars on the MS are represented by the
present-day mass function (PDMF), which
for $1 < m < 100 m_\odot$
is constructed by multiplying the IMF by $m^{-2.5}$ to account for the
lifetime of
massive main sequence stars; for $m<1m_\odot$ the PDMF is
the same as the IMF. A more detailed estimate of the relation between
PDMF and IMF is possible but not needed for our purposes because the  
mass fraction
in stars with masses $m>m_\odot$ is small. We take $m=0.08m_\odot$ as 
the mass dividing hydrogen-burning and substellar mass `brown dwarfs'
(Hayashi \& Nakano 1963; Kumar 1963) and the observational limit
0.01$m_\odot$ as the lower end of substellar masses.

As indicated in the third line of Table 2, we take the average mass of
a white dwarf to be $\langle m\rangle = 0.62\, m_\odot$, for
consistency with the model we describe in section
\ref{sec:stellarbe} (eq.~[\ref{eq:wdmass}]).  This is
close to the value from recent observations, 0.604$\,m_\odot$
(Bergeron \& Holberg, in preparation; Bergeron private
communication). We assume all stars with initial masses in the range
$1<m<8\, m_\odot$ end up as white dwarfs with the adopted average mass,
and the rest of the mass returns to the interstellar medium. With our
adopted IMF this model predicts that the ratio of masses
in white dwarfs to main sequence stars is
$\rho(\hbox{WD})/\rho(\hbox{MS})=0.176$. This can be compared to the
observations. The 2dF survey (Vennes et al. 2002), with the use of
our mean mass $0.62\, m_\odot$, yields a measure of the mass density
of DA white dwarfs,
$\rho(\hbox{DA-WD,local})=(4.2\pm 2.3)\times 10^{-3}
m_\odot\hbox{pc}^{-3}$.
This is multiplied by 1.3 to account for DB, DQ and DZ white
dwarfs (Harris et al. 2003), to give the total white dwarf mass density
$\rho(\hbox{WD,local})=(5.5\pm 3.0)\times 10^{-3}
m_\odot\hbox{ pc}^{-3}$. This divided by the local density of main
sequence stars,
$\rho(\hbox{MS,local})=0.031 \pm 0.002\, m_\odot\hbox{ pc}^{-3}$, from
Reid, Gizis, \& Hawley (2002), and $0.041 \pm 0.003\, m_\odot\hbox{ pc}^{-3}$
from Chabrier (2003),
gives
$\rho(\hbox{WD})/\rho(\hbox{MS})=0.16\pm0.10$,
which  agrees with our model prediction, albeit with a large
uncertainty.

We assume all stars in the initial mass range
\beq
8 < m < 25\,m_\odot\label{eq:msmassrane}
\eeq
end up  as neutron stars with mass $1.35\,m_\odot$ (Nice, Splaver, \&
Stairs 2003; Thorsett \& Chakrabarty 1999), and the rest of the star
mass is recycled. Estimates of
the lower critical mass for stellar core collapse vary from 6 to
10$\,m_\odot$ (Reimers \& Koester 1982; Nomoto 1984; Chiosi, Bertelli,
\& Bressan 1992). The upper critical
mass for the formation of a neutron star is more uncertain; our choice
in equation~(\ref{eq:msmassrane}) is taken from Heger et al. (2003). We
assume all stars with main sequence mass above the limit in
equation~(\ref{eq:msmassrane}) end up as  stellar black hole remnants
with mass $m_\bullet =7.5\,m_\odot$, with the rest of the mass recycled.
This again follows Heger et al. The remnant mass is quite uncertain,
and Heger et al. also indicate that some stars with masses above the
range in equation~(\ref{eq:msmassrane}) may produce neutron stars.

The last column in Table~2
is the cumulative amount of mass that has gone into stars, including
what is later shed, normalised to the mass in column (4). According to
these estimates, 1.6 times the mass in observed stars was used (and
reused) in star formation.

Entries 3.3 to 3.8 in Table~1 are  based on the partition of $\Omega
_{\rm stars}$ among main sequence stars and star remnants in the fourth
column of Table~2, with the FHP partition between the two galaxy
population types in the proportion
\beq
\hbox{spheroid}:\hbox{disc} =0.74:0.26
\label{eq:earlytolate}
\eeq
of mass in ellipticals plus bulges of spirals to the mass in disks plus
irregulars.

\subsubsection{Consistency with the Star Formation Rate}

We can compare our estimate of the mass in stars to  the mass
integrated over the star formation rate density (the SFR) from
high redshift to the present. The H$\alpha$ luminosity density at zero
redshift measured from an SDSS nearby galaxy sample is
${\cal L}({\rm H}\alpha)=10^{39.31{+0.11\atop-0.08}}h$~erg
s$^{-1}$Mpc$^{-3}$ (Nakamura et al. 2004). This agrees with the
earlier value obtained by Gallego et al. (1995).
Within the star formation models Glazebrook  et al. (1999) explored,
1$m_\odot$ yr$^{-1}$ of star
formation produces H$\alpha$ luminosity
$L_\alpha=\big(2.00{+0.92\atop-0.24}\big)\times 10^{41}$ erg s$^{-1}$
for
our adopted IMF. The ratio of ${\cal L}({\rm H}\alpha)$ to $L_\alpha$
is an estimate of the present-day  SFR,
\beq
\psi (0)=0.0071^{+0.0021}_{-0.0035}\,m_\odot~ {\rm yr}^{-1}{\rm
Mpc^{-3}}.
\label{eq:localsfr}
\eeq

The evolution of the SFR at $z\la 1$ is now observationally
well determined (Lilly et al. 1996; Heavens et al. 2004; see Glazebrook
et al. 1999 for a summary of the measurements up to that time), and may
be approximated as
\beq
\psi (t) = \psi (0) e^{(t_o - t)/\tau_{\rm sf}}, \quad z < 0.85,
\label{eq:sfrhistory}
\eeq
where
\beq
     \tau_{\rm sf}\simeq 2.9\hbox{ Gyr},
\eeq
and $t_o - t$ is the time measured back from the present in the
$\Lambda$CDM cosmology (Fukugita \&
Kawasaki 2003; see also
Glazebrook et al. 1999). The  situation at higher redshift is still
controversial (Madau et al. 1996; Steidel et al. 1999). We shall
suppose that the SFR is
constant from $z=0.85$ back to the start of star formation at redshift
$z_f$, at the value
\beq
\psi = 0.0877 \, m_\odot~ {\rm yr}^{-1}{\rm Mpc^{-3}},
     \ 0.85 < z < z_f.\label{eq:sfrearly}
\eeq
In this model the integrated star formation is
\beqa
\Omega _{\rm stars} &\simeq & 0.0038\hbox{  if  }z_f=2,\nonumber\\
&\simeq & 0.0047\hbox{  if  }z_f=3,\label{eq:halphasfr}\\
&\simeq & 0.0053\hbox{  if  }z_f=5.\nonumber
\eeqa
There is a significant downward uncertainty,
$\delta\Omega _{\rm stars} \sim -0.0025$, arising from the uncertainty
in the star formation rate (eq.[\ref{eq:localsfr}]).
Equation~(\ref{eq:omegastars}) and Table~2
indicate that the mass processed into stars, including what was later
shed, is $\Omega=0.0027\times 1.62=0.0044$. This is consistent
with the cumulative star formation in equation~(\ref{eq:halphasfr}).
There would be a problem with the numbers if there were reason to
believe that the SFR continued to increase with increasing redshift
well beyond $z=1$. A constant or declining SFR at $z>1$ is well
accommodated with our estimates for the baryon budget and current ideas
about the star formation history at high redshift.

The star formation history determines the average reciprocal redshift
factor (as in eq.~[\ref{eq:integraloflumden}]) for the effect of
redshift on the integrated comoving energy density of
electromagnetic radiation (and other forms of relativistic mass)
generated by stars.  In our model for the SFR the correction factor is
\beqa
\langle (1+z)^{-1}\rangle ^{-1} &=& {\int dt\, {\cal L}\over\int dt\,
{\cal L}/(1+z)} \nonumber\\
&=& 2.0\pm0.15,\label{eq:redshiftloss}
\eeqa
where ${\cal  L}(t)$ is the bolometric luminosity density per comoving
volume, which we are assuming is proportional to the star formation
rate density.  The numerical value assumes the redshift cutoff is in the
range $z_f=2$ to~5. For the purpose of comparison of the accumulation
of stellar products to the present rate of production, another useful
quantity represents the integrated comoving density  of energy radiated
in terms of the effective time span
normalised by the present-day luminosity density,
\beq
\Delta t_{\rm eff}={\cal L}(t_0)^{-1}\int dt {\cal L}(t)
= 85\pm13\hbox{ Gyr},
\label{eq:timespan}
\eeq
The value assume the range of models for the star formation history in
equation~(\ref{eq:halphasfr}).

We can compare the rate of stellar core collapse in our model to
observations of the supernova rate. (See Fukugita \& Kawasaki 2003 and
Madau, della Valle \& Panagia 1998 for similar analyses.)
The present SFR in
equation~(\ref{eq:localsfr}) and the critical minimum mass
8$\,m_\odot$  in equation~(\ref{eq:msmassrane}) imply that the present
rate of formation of neutron stars and stellar black holes
is expected to be
\begin{eqnarray}
R &=&
                 {\int^{100}_8 dm~ dN/dm\over \int^{100}_{0.08}
dm~m~dN/dm}\nonumber\\
         &=& 0.0079^{+0.0024}_{-0.0039}~ (100{\rm yr})^{-1}\hbox{\rm
Mpc}^{-3}.\qquad
\end{eqnarray}
Our estimate of the observed supernova rate is
\beq
R_{\rm SN}=0.0076^{+0.0064}_{-0.0020}~
(100{\rm yr})^{-1}\hbox{\rm Mpc}^{-3}.\label{eq:obssnerate}
\eeq
This is the geometrical mean of the rates from three surveys for
Type II and Type Ib/c supernovae, 0.037, 0.018 and 0.017, in units of
$h^3(100{\rm yr})^{-1}{\rm Mpc}^{-3}$ (Tammann, L\"offler \& Schr\"oder
1994;
Cappellaro et al. 1997; van den Bergh \& McClure 1994; see also
Cappellaro, Evans \& Turatto 1999). We conclude that, if most stars with
initial masses greater than about $8\,m_\odot$ produced Type II and
Ib/c supernovae,
then our model for the star formation history would pass this
consistency check.

We remark that in our model the comoving number density of supernovae
of types II and Ib/c integrated back to the start of star formation is
\beq
R \Delta t_{\rm eff} = 7\pm 3\times 10^6\hbox{ Mpc}^{-3},
\label{eq:netsneproduction}
\eeq
where the effective time span is given in equation (\ref{eq:timespan}).

\subsubsection{Planets and Condensed Matter}

The mass in planets that are gravitationally bound to stars must be
small, but
it is of particular interest to us as residents of a planet. Marcy
(private communication; see also Marcy \& Butler 2000) finds that about
6.5\%\ of nearby FGKM stars have
detected Jovian-like planets, and that an
extrapolation to planets at larger orbital radii might be expected to
roughly double this number. In our model for the PDMF the ratio of the
number density of stars in the mass range 0.08 to $1.6~m_\odot$ to the
mass density in stars is $n/\rho = 2.1m_\odot ^{-1}$. The product of
this quantity with the mass density in stars
(eq.~[\ref{eq:omegastars}]), the fraction 0.13, and the ratio of the
mass of Jupiter to the Solar mass is
\beq
\Omega _{\rm planets} = 10^{-6.1}.
\label{eq:planets}
\eeq
Marcy indicates that stars with
lower metallicity have fewer planets, but that may not introduce a
serious error because there are fewer low metallicity stars.

There is a
population of interstellar planets that have escaped from stars that
have rapidly shed considerable mass. If the number of planets per star 
is independent of the stellar mass then, in our IMF, stars in the mass 
range
1 to 1.6$m_\odot$ had 0.025 times the number of stars that are bound to 
main sequence stars. Among them about 2/3 are likely to be
swallowed by the host stars in their giant star phases. This leaves
the mass in liberated planets, and those still bound to white dwarfs, 
at about 0.07 times the mass density in equation~(\ref{eq:planets}).
Planets associated with more massive stars may more likely escape 
during core collapse, but the numbers of stars, and perhaps planets, is 
small. The large uncertainty in equation~(\ref{eq:planets}) is
whether the stellar neighborhood is a fair sample. This leads us to 
enter
the order of magnitude in Table~1.

We consider separately the mass in objects small enough to be held
together by molecular binding energy rather than gravity. The dominant
amount of the former is interstellar dust. It is known (Draine 2003)
that $\ga 90$\%
of silicon atoms in interstellar matter are condensed into grains,
probably
dominantly in the forms of enstatite (MgSiO$_3$) and forsterite
(Mg$_2$SiO$_4$) with the former likely three to four times
more abundant (Molster et al. 2002). This means the mass in this form
is  $Z(\hbox{silicates})/Z=0.17$ times the mass in heavy elements.
Draine (2003) suspects that the dominant contribution to the
carbonaceous material is in the form of polycyclic aromatic
hydrocarbon with the inferred abundance C/H=6$\times10^{-5}$.
Adding this to the silicates, we find that the mass fraction
becomes $Z(\hbox{dust})/Z=0.20$. The formation of dust with iron or
other forms of carbonates could increase this number. Entry 3.12,
$\Omega_{\rm dust}=10^{-5.6}$, is the
product of the density parameter in cool gas (entries 3.9 plus 3.10)
with the mean
metallicity discussed below (and displayed as
eq.~[\ref{eq:meanmetallicity}]) and the 20 percent heavy element mass
fraction in dust. Our estimate, which assumes that what we know about
the Milky Way applies to other galaxies, is crude, but it seems likely
that the mass in dust exceeds the mass in planets.

There are larger objects in which molecular binding is important. The
gravitational binding energy of a roughly spherical object with mass
$m$ and radius $r$ is $\sim Gm^2/r$, and the molecular binding energy
is roughly 1 eV per atom. The  molecular binding energy is larger than
the
gravitational energy when
\beq
m\la 1.2\times 10^{28}~\hbox{ g}~(5\hbox{ g cm}^{-3}/\rho)^{1/2}/\bar A,
\eeq
where the mass density is $\rho$ and the mean atomic weight is $\bar
A$. For silicates, this bound is $m=8\times 10^{26}$ g. That is,
gravitational binding energy dominates in the Earth and molecular
binding energy dominates in the Moon and asteroids. Trujillo, Jewitt,
\& Luu (2001) estimate that the mass in the Kuiper Belt objects is
about a tenth of an Earth mass, or $10^{-6.5}m_\odot$. A comparable
mass is in the moons in the Solar system. If this mass fraction were
common to all stars the density parameter in these objects would be
\beq
\Omega _{\rm asteroid}\sim 10^{-9},\label{eq:asteroid}
\eeq
a small fraction of the mass in dust.

\subsubsection{Neutral Gas}

The recent blind HI surveys are a significant advance over the data
used by FHP to estimate the mass density in neutral atomic gas. The
largest survey, HIPASS, with 1000 galaxies (Zwaan et al. 2003),
gives
\beq
\Omega_{\rm HI}=4.2\pm 0.7\times 10^{-4}.
\eeq
The increase over the value quoted in FHP (1.5 times the upper
end value of FHP) illustrates the
advantage of blind surveys over observations of programmed galaxies.
The molecular hydrogen abundance from the CO survey of Keres, Yun \&
Young (2003) is
\beq
\Omega_{\rm H_2}={1.6\pm0.6}\times 10^{-4}.
\eeq
The sum of these two values is multiplied by 1.38 to
accommodate helium.

We place in entry 3.9 the atomic hydrogen and the helium abundance
belonging to atomic and molecular hydrogen. Entry 3.10 is the molecular
hydrogen component. The mass in this neutral gas is 1.7$\pm$0.4\% of the
total baryon mass.

\subsubsection{Intracluster Plasma}

The estimate of the plasma mass in rich clusters of galaxies depends on
a
convention for the cluster radii and masses. We use the mass $M_{200}$
contained by the nominal virial radius, $r_{200}$.
This definition is more faithful to a physical definition of the part
of a cluster that is close to dynamical equilibrium, and it also traces
the X-ray radius which is the definition of our hot gas. In the
limiting isothermal sphere model the relation to the mass within the
Abell
radius $r_A=1.5h^{-1}$~Mpc is
\beq
M_A = M_o^{1/3}M_{200}^{2/3}, \quad M_o=1.1\times 10^{15}m_\odot .
\eeq
This agrees with the estimates of the two
masses given by Reiprich \& B{\"
o}hringer (2002),  Table~4. We adopt the Reiprich \& B{\" o}hringer
cluster mass density parameter,
\beq\Omega _{\rm cl} = 0.012^{+0.03}_{-0.04},\label{eq:omegacluster}\eeq
for the cluster mass limit
\beq
M_{200}>5\times 10^{13}m_\odot, \quad
M_A>1.3\times 10^{14} m_\odot .\label{eq:clusteromega}
\eeq
That is, 4\% of the mass is assembled in rich clusters.
The Bahcall \&\ Cen (1993) Abell mass function is consistent with the
Reiprich \& B{\" o}hringer estimate. We caution that the
Bahcall et al. (2003)  SDSS mass function is about half what we adopt,
perhaps because SDSS samples a subset of the clusters.
Note also that the integral over the cluster mass within the Abell
radius
gives
a substantially larger value for $\Omega _{\rm cl}$ because $M_A>M_o$
at the masses $M\ga 10^{14}m_\odot$ that dominate the integral for a
given mass function.

The abundance of hot baryons in clusters is obtained from the expression
\begin{eqnarray}
\Omega_{\rm b, cl} &=&0.012
(\Omega_{\rm b}-\Omega_{\rm stars})/\Omega_m\nonumber\\
&=&0.0018\pm0.0007.
\end{eqnarray}
The change from the value in FHP is mainly due to the definition
of the cluster mass, as also noted by Reiprich \& B\"ohringer.

\subsubsection{Massive Black holes}

We follow the standard idea that the massive objects in the centers of
galaxies are black holes that formed by the accretion of baryons.
Baryons entering black holes are said to lose their identity, but for
our accounting it is appropriate to consider them to be sequestered
baryons.

We define the characteristic efficiency factor $\epsilon _n$ for black
hole formation (where the subscript is meant to distinguish the massive
black holes in the centers of galaxies from stellar mass black holes)
by the black hole mass produced out of an initial diffuse baryon mass
$m_{\rm b}$,
\beq
m_{\rm bh}=(1-\epsilon _n)m_{\rm b},\label{eq:bhbe1}
\eeq
where the energy released in electromagnetic radiation, neutrinos,
kinetic energy, and possibly gravitatonal radiation is
\beq
m_{\rm emitted} = \epsilon _nm_{\rm b}.\label{eq:bhbe2}
\eeq
The baryon mass sequestered in massive black holes, $m_{\rm b} = m_{\rm
bh}/(1-\epsilon _n)$, could be substantial if $\epsilon _n$ were
close to unity. However, if an appreciable part of the
binding energy were released as electromagnetic radiation then the
bounds on the radiation background (in category 7) and
those of the CMBR distortion  (eq.~[\ref{eq:y-parameter}])
would require that
the energy was released at very high redshift, $kT>10^7$~K.
The estimates discussed in \S 2.7.6 indicate the mass fraction  
released in  gravitational radiation is small. Thus it seems likely 
that $\epsilon _n$ is small, as is assumed in
entry 3.13. The efficiency factor $\epsilon _s$ typical of stellar mass
black holes does not appear in entry 3.7, because this estimate is
based on an
analysis of the progenitor star masses. The estimate of the mean mass
density in massive black holes that is used for entry 3.13 is discussed
in \S 2.5.3.

\subsubsection{Intergalactic Plasma}

Entry 3.1, for the baryon mass outside galaxies and clusters of
galaxies, is the difference between our adopted value of the baryon
density
parameter (eq.~[\ref{eq:baryonomega}]) and the sum of all the other
entries in category 3. Within standard pictures of structure formation
this component could not be in a compact form such as planets,  but
rather must be a plasma, diffuse enough to be ionized by the
intergalactic
radiation or else shocked to a temperature high enough for collisional
ionization, but not dense and hot enough to be a detectable X-ray
source outside clusters and hot groups of galaxies.

In our baryon budget 90\% of the baryons are in this intergalactic
plasma. It is observed in several states. Quasar
absorption lines show matter in  low and high atomic ionization states
in the halos of $L\sim L_\ast$ galaxies, extending to radii $\sim
200$~kpc
(Chen et al. 2001 and references therein). Absorption lines of HI and
MgII reveal low surface density photoionized plasma at kinetic
temperature $T\sim 10^4$~K, which can be well away from $L\sim L_\ast$
galaxies, as discussed by Churchill, Vogt, \& Charlton (2003), and
Penton, Stocke \&\ Shull (2004). The latter authors estimate that 30\%
of
the baryons are in this state. Absorption lines of O~VI around galaxies
and groups of galaxies (Tripp, Savage \& Jenkins 2000; Sembach et al.
2003; Shull,
Tumlinson, \& Giroux 2003; Richter et al. 2003), reveal matter that may
be excited by photoionization by the X ray background radiation and by
collisions in plasma at the kinetic temperature $T\sim 10^6$~K
characteristic of the motion of matter around galaxies
(Cen et al. 2001). The detection of  O~VII and O~VIII absorption lines 
also indicates the presence of higher temperature regions in the local 
IGM (Fang et al. 2002; Fang, Sembach \& Canizares 2003).
Improvements in the constraints on the amount of
matter in these various states of intergalactic baryons  will be
followed with interest.

For the inventory we adopt the measure of the concentration of dark
matter around galaxies in equation~(\ref{eq:galaxydarkmass}), and the
argument discussed in \S~2.3.8 that the baryons are distributed like
the dark matter on scales comparable to the virial radii of galaxies.
The resulting division into the mass in baryons near the virial radii
of normal galaxies outside clusters (entry 3.1a), and the mass well
away from galaxies and compact groups and clusters of galaxies (entry
3.1b), is presented as sub-components, because the sum is much better
constrained than the individual values.

\subsubsection{Baryon Cooling}

We comment here on a simple picture for the cooling and settling of
baryons onto galaxies.
The sum of the baryon mass densities belonging to galaxies, in entries
3.3 to 3.13, is $\Omega _{\rm b, g}=0.0035$. This is 8\% of
the total baryon mass. Suppose $\Omega _{\rm b, g}$ consists of all
baryons gathered from radius $r_g$ around $L\sim L_\ast$ galaxies,
and suppose we can neglect the addition of baryons by settling from further
out and the loss by galactic winds. That is, we are supposing that at
$r>r_g$ the ratio of the baryon density to the dark matter density is
the cosmic mean value, and that the baryons closer in have collapsed
onto the galaxies. In this picture the characteristic radius of
assembly of the baryons satisfies
\beq
{2 n_g r_g \sigma^2 \over G \rho_m}={\Omega _{\rm b, g}\over
\Omega _{\rm b, total}}, \label{eq:xxxx}
\eeq
in the limiting isothermal sphere approximation
(eq.~[\ref{eq:isothermal}]).

In equation~(\ref{eq:xxxx}) $n_g$ is a measure of the number density of
luminous galaxies. We record here our choices for this quantity and
related parameters that are used elsewhere. We take
\beq
n_g={\cal L}_r/L_r^*=0.017h^{3}~{\rm Mpc}^{-3},
\label{eq:galdensity}
\eeq
where ${\cal L}_r$ is the luminosity density
(eq.~[\ref{eq:luminositydensity}]). The characteristic galaxy
luminosity,
\beqa
&&L_B^\ast = 1.07\times 10^{10}h^{-2} L_\odot,\nonumber\\
&&L_r^\ast = 1.45\times 10^{10}h^{-2} L_\odot ,\label{eq:charlum}\\
&&L_z^\ast = 2.37\times 10^{10}h^{-2} L_\odot ,\nonumber
\eeqa
is the luminosity parameter in the Schechter function, with the power
law index $\alpha _B=-1.1$, $\alpha _r =-1.13$,  and $\alpha _z=-1.14$.
We refer some estimates of energy densities to what is known about the
Milky Way, for which we need the effective number density of Milky Way
galaxies. In the B band the Milky Way luminosity is $L_{\rm MW} =
1.3L_B^\ast$, and the effective number density is
\beq
n_{\rm MW}={\cal L}_B/L_{\rm MW}^*=0.013h^{3}~({\rm Mpc})^{-3}.
\label{eq:MWgaldensity}
\eeq
Almost the same density follows when referred to the r-band. The Milky 
Way parameters are discussed in the Appendix.

With the characteristic velocity
dispersion in equation~(\ref{eq:virial}) and the characteristic galaxy
number density in equation~(\ref{eq:galdensity}) the baryon accretion
radius
defined by equation~(\ref{eq:xxxx}) is
\beq
r_g\simeq 30h^{-1}\hbox{ kpc},\label{eq:accrretionradius}
\eeq
at density contrast $1.3\times 10^4$ and plasma
density $n_{\rm gas}(r_g)\sim 0.007h^3$~cm$^{-3}$.
If plasma at this radius were supported by pressure at the
one-dimensional velocity dispersion
$\sigma\simeq 160$ km~s$^{-1}$ in equation~(\ref{eq:virial})
the temperature would be $T\simeq 2\times 10^6$~K. At this
density and temperature the thermal bremsstrahlung cooling time would be
short enough, $\sim 4\times10^9$ yr, that stars would have formed
and disks matured at $z\sim 1$.

We have lower bounds on the cooling radius from the observation that
the neutral atomic hydrogen density around $L_*$ galaxies
reaches $N_{\rm HI}=1.8\times 10^{20}$ cm$^{-2}$ at the
effective radius $\sim 20h^{-1}$ kpc (Bosma 1981),
and Mg II absorption
lines are observed at radius $\sim 40h^{-1}$ kpc (Steidel,
Dickinson \& Persson 1994).
For the Milky Way the distribution of RR Lyr stars cuts off sharply at
50 kpc (Ivezi\'c et al. 2000).
The cooling radius must be larger than these indicators of relatively
cool matter. Equation~(\ref{eq:accrretionradius}) is not inconsistent
with this condition.
Though the history of baryon accretion
by galaxies undoubtedly is complex, we can imagine, as a first
approximation, that a substantial fraction of the baryons now
concentrated in galaxies are there because they were able to cool
and settle from an initial distribution similar to that of the
dark matter.

The relative distributions of baryons and dark matter at distances much
larger than $r_g$ from galaxies might not be greatly disturbed from the
primeval condition. If so, then the product of the baryon density
parameter with the virialized dark matter mass fraction in
equation~(\ref{eq:galaxydarkmass}) is an estimate of the baryon mass
that resides within the virial radii of normal galaxies, and the
remainder,
\beq
\Omega _{\rm b, ig}\sim 0.016,\label{eq:igmfraction}
\eeq
which is presented as entry 3.1b, would be located outside galaxies and
remain less than fully documented.

\subsection{Primeval Gravitational Energy}

In the $\Lambda$CDM cosmology the gravitational binding energy of the
present mass distribution has two contributions. The first, which we
term primeval, is a result of the purely gravitational growth of mass
fluctuations out of the small adiabatic departures from a homogeneous
mass distribution present in the initial conditions for the
Friedmann-Lema\^\i tre cosmology. The second, to be discussed in the
next subsection, is the result of dissipative settling of baryons that
produced
the baryon-dominated luminous parts of the galaxies along with stars
and star remnants. We can find sensible approximations to the
primeval and dissipative components because, as we will discuss, the
characteristic length scales are well separated.

The primeval gravitational energy is defined by imagining a universe
with initial conditions identical to ours in all respects except that
the
baryonic matter in our universe is replaced by an equal mass of CDM in
the reference model. At the present world time this reference model
contains a clustered distribution of massive halos with gravitational
binding energy density that we term the primeval
component.\footnote{One surely would say that in
this model universe the virialized dark matter halos have gravitational
binding
energy. Since there was no energy transfer to some other form, one
might also want to say that this binding energy must have been present
in the initial conditions. Furthermore, one can assign to a linear mass
density fluctuation with contrast $\delta(t) > 0$, and comoving radius
$x$ (physical radius $xa(t)$), a gravitational energy per unit mass,
$W' \sim  - G \langle\rho\rangle\delta (ax)^2$, which is constant in
linear perturbation theory and comparable to the binding energy of the
final virialized halo. Perhaps one can use this as a guide to a
definition of the primeval energy belonging to the density fluctuation,
despite the problem that the mean of $W'$ vanishes, and the fact that
in general relativity theory there is no general definition of the
global energy density of a statistically homogeneous system. We have
not been able to find a useful approach along these lines. We
might add that if our universe had been Einstein-de Sitter then we
would have defined the primeval energy density by a numerical solution
of equation~(\ref{eq:LIrvine}).} This component is
estimated as follows.

The Layzer (1963) - Irvine (1961) equation for the evolution of the
kinetic and gravitational energies of nonrelativistic matter, such as
CDM, that interacts only by gravity is
\beq
{d\over dt} (K + W)  + {\dot a\over a}(2K + W) = 0.\label{eq:LIrvine}
\eeq
The kinetic energy per unit mass is
\beq
K=\langle m\vec v^{\,2}/2\rangle /\langle m\rangle,
\eeq
where  $\vec v$ is the peculiar velocity of a particle with mass $m$.
The gravitational potential energy per unit mass is
\beq
W = - {1\over 2} G \rho _m\int d^3r\,\xi (r)/r ,
\eeq
where $\rho _m$ is the mean mass density and $\xi$ is the reduced mass
autocorrelation function.

We can use a simple limiting case of the Layzer-Irvine equation,
because in the $\Lambda$CDM cosmology the universe has now entered
$\Lambda$-dominated expansion, which has caused a significant
suppression of the rate of growth of large-scale departures from homogeneity.
This means that the first term in equation~(\ref{eq:LIrvine}) has become
small compared to the second term.
Thus it is a reasonable approximation to take $2K=-W$, the usual virial
equilibrium relation. Then the gravitational binding energy per unit
mass is $U = K + W = W/2$, and the cosmic mean primeval gravitational
binding energy is the product of $U$ with the mean mass density $\Omega
_m$ in matter (eq.~[\ref{eq:omegam}]). With the normalization
$P(k)=\int d^3r\,\xi (r)e^{i\vec k\cdot\vec r}$ for the mass
fluctuation power spectrum, the expression for the primeval
gravitational binding energy in this approximation is
\beq
\Omega _{\rm BE, p} = -{3\Omega _m^2H_o^2\over 16\pi ^2}\int _o^\infty
dk\,
P(k).
\label{eq:omega5}
\eeq

For a numerical value we use the mass fluctuation power spectrum $P(k)$
in Figure 37 of Tegmark et al. (2004b) at $k<0.1h$~Mpc$^{-1}$. At
smaller scales the Tegmark et al. spectrum decreases too rapidly with
increasing wavenumber $k$ to be a good approximation to the present
mass distribution (Davis \& Peebles 1983).
We approximate the spectrum as
\beq
P(k) = 7000 \left( 0.1h\hbox{ Mpc}^{-1}\over k\right)
^{1.23}h^{-3}\hbox{ Mpc}^{3},
\label{eq:smallscalepower}
\eeq
at $k>0.1h$~Mpc$^{-1}$. This is based on the Fourier transform of the
pure power law model for the galaxy autocorrelation function
(eq.~[\ref{eq:galaxyxi}]),
\beq
P = 4\pi r_o^\gamma k^{\gamma - 3}\Gamma (2 - \gamma )
\sin\pi (2 - \gamma )/2,
\eeq
which gives $P=5200h^{-3}$~Mpc$^3$ at $k=0.1h$~Mpc$^{-1}$. We choose
the somewhat larger normalization in
equation~(\ref{eq:smallscalepower}) to fit the SDSS measurement. The
numerical result is
\beq
\int _o^\infty P(k)\, dk = 4800 h^{-2}\hbox{ Mpc}^2.
\label{eq:pofk}
\eeq
The integral of the Tegmark et al. (2004b) power spectrum over all
wavenumbers is $2700h^{-2}$Mpc, which is
0.55 times the value in  equation~(\ref{eq:pofk}). Our
estimate of the integral is based on measurements of the actual power
spectrum, not the
spectrum that would have obtained if there were no dissipative settling
of baryons, but the error is small because the integral is
dominated by lengths large compared to the scale of separation of
baryons from the dark matter.

Equations~(\ref{eq:omega5}) and~(\ref{eq:pofk}) yield the entry in line
4 of Table~1,
\beq
\Omega_{\rm grav, p}=-7.7\times10^{-7}. \label{eq:primevalbe}
\eeq
This is the density parameter of the gravitational
binding energy of the present departure from a homogeneous mass
distribution, ignoring the effects of the dissipative settling of
baryons.
It will be useful to note
that a measure of the length scale of the gravitational energy is the
half-point of the integral in equation~(\ref{eq:pofk}), at wavenumber
$k_{1/2}=0.27 h$~Mpc$^{-1}$, or half wavelength
\beq
\lambda _{1/2}=\pi /k_{1/2}=12h^{-1}\hbox{ Mpc}.
\label{eq:halflambda}
\eeq

The kinetic energy per unit mass belonging to
equation~(\ref{eq:primevalbe}) is
\beq K = - U = 3(410\hbox{ km s}^{-1})^2/2.\eeq
The velocity in parentheses is the one-dimensional single-particle
line-of sight rms peculiar velocity.

The gravitational binding energy is not equal to a sum over the
contributions from individual objects, but we can write useful
approximations to the decomposition into the three components --- the
virialized parts of the massive halos of $L\sim L_\ast$ galaxies, the
rich clusters, and large-scale clustering --- shown in category 4 in the
inventory.

Since galaxy rotation curves tend to be close to flat we write the
binding energy of the virialized parts of the dark matter halos of the
galaxies as
\beq
\Omega _{\rm grav}  = -{3\Omega _m\sigma ^2\over 2}
{\rho (r_v)\over \rho _m }=10^{-7.2},\label{eq:omegagrav}
\eeq
where $\Omega _m$ is the matter density parameter
(eq.~[\ref{eq:omegam}]),  $\sigma$ is the characteristic velocity
dispersion in
equation~(\ref{eq:virial}) and the last factor is the virialized mass
fraction (eq.~\ref{eq:galaxydarkmass}).  This is about 10\%\ of the
total gravitational binding energy. The dissipative settling that
produced the
baryon-dominated luminous parts of the galaxies would have perturbed
the massive halos, but the disturbance to the primeval gravitational
energy is small because the baryon mass fraction is small.

Our estimate of the binding energy of the dark halos of the galaxies
may be compared to the density parameter given by
equation~(\ref{eq:omega5}) when the integral over $P(k)$ is restricted
to small scales, $k>\pi /r_v$, where $r_v$ is the virial radius in  in
equations~(\ref{eq:virial}) and~(\ref{eq:omegagrav}). The result, with
equation~(\ref{eq:smallscalepower}) for $P(k)$, is twice the value in
equation~(\ref{eq:omegagrav}). The difference is an indication of the
ambiguity of separating gravitational energy into components. For entry
4.1 we adopt the value in equation~(\ref{eq:omegagrav}) as the more
directly interpretable.

Equation (\ref{eq:omegagrav}) is a reasonable approximation for most of
the mass in galaxy-size dark halos in luminous field galaxies such as
the Milky Way, but in rich clusters the galaxies tend to
share a dark halo that is close to smoothly distributed across the
cluster. Our estimate in entry~4.2 for the  primeval gravitational
binding energy belonging to rich clusters follows
equation~(\ref{eq:omegagrav}), with $\sigma = 800$ km~s$^{-1}$ and
$\Omega _{\rm cl}$ from equation~(\ref{eq:omegacluster}).
Rich clusters share about 15\% of the total gravitational binding
energy.

Entry 4.3 is the difference between the total in entry 4 and the sum of
entries 4.1 and 4.2. The Layzer-Irvine equation indicates that the
binding energy is dominated by a length scale
(eq.~[\ref{eq:halflambda}]) that is much larger than galaxy virial
radii. Consistent with this, entry 4.3 is larger than entry 4.1 for the
binding
energy of the dark halos of galaxies. The difference is not
large, however, because at small scales the integral over the power
spectrum converges slowly, as $k^{-0.23}$.

\subsection{Dissipative Gravitational Settling}

Dissipative settling has increased the magnitude of the gravitational
binding energy from that prescribed by the primeval conditions
considered in the last section.  In \S 2.5.1 we discuss the energy
released in
producing the increased mean density of baryons relative to dark matter
in the luminous parts of the galaxies, in \S 2.5.2 we
estimate the gravitational energy released in stellar formation and
evolution, and in \S 2.5.3 we consider the central massive
compact objects in galaxies.

\subsubsection{The Luminous Parts of Galaxies}

In the Milky Way galaxy the mass within our position, at about 8~kpc
   from the center, is roughly equal parts baryonic and dark matter,
or about 6 times the cosmic mean ratio (eq.~[\ref{eq:omoverob}]). This
is thought to be the usual situation in the luminous parts of normal
galaxies. The amount of gravitational binding energy released in
producing this concentration of baryons depends on how it was done. In
one limiting case one may imagine that stars formed in the
centers of low mass dark halos with relatively small dissipation of
energy --- apart from star formation --- because the depths of the
gravitational potential wells were small, and that the low mass halos
later
merged without any additional dissipation, the dense baryon-dominated
parts remaining near the densest regions to form the present-day
baryon-dominated luminous parts of galaxies. (This is an extreme
version of the scenario discussed by Gao et al. 2003). In another
extreme, one
may imagine that the baryons settled into previously assembled
galaxy-scale halos, which would dissipate considerably more energy. A
galaxy has
a definite computable gravitational binding energy, of course (apart
   from the difficulty of correcting for ongoing accretion), but to 
relate
this to the energy dissipated in producing the galaxy would require an
analysis of what the mass distribution would have been in the absence
of dissipation, which is not an easy task.

These considerations lead us to offer only a crude
estimate for entry 5.1, which we write as the product of the density
parameter belonging  to baryons in galaxies --- the sum $\Omega _{\rm b,
g} =0.0035$ of
the density parameters in entries 3.3 to 3.13 --- with the kinetic
energy per
unit mass, $K = 3\sigma ^2/2$ and $\sigma =160$ km~s$^{-1}$. The result
is a  2\% addition to the primeval halo gravitational binding energy
(entry 4.1). If the baryon concentrations in galaxies formed at high
redshifts in small
halos the dissipative energy released would be an even smaller fraction
of the total.

\subsubsection{Stellar Binding Energy}
\label{sec:stellarbe}

The amount of binding energy released in star formation is easy to
define
because the relative length scale is small. We write the gravitational
binding energy per unit mass for a star with mass $m$ and radius $r$ as
\beq
{{\rm BE}\over m} = - K {Gm\over rc^2}.
\eeq
The prefactor for the Sun is $K_\odot=1.74$, and
$K=0.3$ for a homogeneous sphere.

For main sequence stars we use  the zero-age mass-radius
relation, $r\simeq 0.85\, m^{0.80}$ for $0.08<m<0.79$,
$r\simeq 0.93\, m^{1.17}$ for $0.79<m<1.38$, and
$r\simeq 1.15\, m^{0.52}$ for $1.38<m<100$, in solar units. These
numbers are assembled from Ezer \& Cameron (1967), Cox \& Giulli (1968)
and
Cox (2000). Integration of $Gm^2/r$ over the PDMF gives
${\rm BE}/m=3.7\times 10^{-6}$. The product of the last number with
the density parameter of the mass in main sequence stars (entries 3.3
plus 3.4), with $K\simeq K_\odot$, is the estimate of the gravitational
binding energy,  $\Omega_{\rm BE}=-10^{-8.1}$, for stars. We similarly
obtain the substellar gravitational binding energy, $\Omega_{\rm
BE}=-10^{-9.6}$, where $r$ is fixed at $0.096\, r_\odot$ (Burrows et al.
2001). This is a small addition to the
sum in entry 5.2.

We construct a model for the white dwarf mass function from an
approximation to the relation between the progenitor main sequence mass
and the white dwarf remnant mass (Claver et al. 2001; Weidemann 2000),
\beq
m_{\rm wd}=0.08 m_{\rm ms}+0.45\, m_\odot ,
\label{eq:mwd-mms}
\eeq
and our IMF. White dwarf masses run from 0.53 to 1.09\,$m_\odot$
for the main sequence mass range $1<m_{\rm ms}<8\,m_\odot$ we have
adopted. The white dwarf mass function
$dN/dm_{\rm wd}=(dN/dm_{\rm ms})(dm_{\rm ms}/dm_{\rm wd})$
thus obtained
agrees well with the observed white dwarf mass distribution
of Bergeron \& Holberg for $m\ga 0.5m_\odot$
(in preparation). Here we ignore low mass helium core white dwarfs.
In our mass function
the mean white dwarf mass is
\beq
\langle m _{\rm wd}\rangle=0.62\,m_\odot.\label{eq:wdmass}
\eeq
   From our mass function and the mass-radius relation given
by Shapiro and Teukolsky (1983) we obtain the mean white dwarf
gravitational binding energy per unit mass,
\beq
{{\rm BE}\over m}=-58 K_{\rm wd}{Gm_\odot\over r_\odot c^2}=-1.2\times
10^{-4}.
\eeq
Since the fractional half-mass radius of a white dwarf is 0.57
times the solar value (Schwarzschild 1958), we have taken
$K_{\rm wd}\simeq 1.0$.
The product with the mass density in white dwarfs (entry 3.5) is entry
5.3.

We take the binding energy of a neutron star to be
$3\times 10^{53}$ erg (e.g., Burrows 1990; Janka \& Hillebrandt 1989),
or
\beq
{\rm BE}/m = -0.12.
\eeq
The product with entry 3.6 is entry 5.4,
$\Omega_{\rm BE,NS}=10^{-5.2}$.
The gravitational binding energies in neutron stars and stellar mass 
black holes are substantially larger than the gravitational binding 
energies in
all other forms.

\subsubsection{Black Hole Binding Energy}

Our definition of the binding energy associated with a black hole
requires careful explanation because it has some curious properties,
including violation
of the thought that it would be logical to consider the mass of a black
hole to be purely gravitational if the matter out of which it formed
has lost its existence.

We choose the definition by analogy to
nuclear and Newtonian gravitational binding energy, in terms of the
energy liberated in the assembly of a system out of its initial parts,
that is, the difference between the total mass of the initial parts and
the mass of the assembled system. In the same way, we use
equations~(\ref{eq:bhbe1}) and~(\ref{eq:bhbe2}) to define the binding
energy of a black hole by the difference between the mass $m_{\rm b}$
of the initial parts --- baryons --- and the mass $m_{\rm bh} = (1 -
\epsilon
_n)m_{\rm b}$ of the final black hole. Thus our definition of the
binding energy of a black hole is
\beq
{\rm BE} = -\epsilon _nm_{\rm b}= -{\epsilon _n\over 1-\epsilon _n}
m_{\rm  bh}.
\label{eq:bhbindingenergy}
\eeq
The magnitude of BE is the energy emitted as electromagnetic and
gravitational
radiation, neutrinos, and kinetic energy, as is appropriate for our
purpose of telling the energy transfers and balancing the baryon budget.
It will be noted that in this definition the binding energy depends on
how the black hole formed. For example, a Solar mass black hole that
formed with
efficiency $\epsilon=0.99$ is assigned binding energy $-99\, m_\odot$,
because it released that much energy, while an identical black hole
that formed with $\epsilon _n=0.01$ is assigned a very different
binding energy, $-0.01\, m_\odot$.

Entry 5.5 for the gravitational binding energy of stellar mass black
holes is the product of entry 3.7, which is our estimate of the
baryonic mass entering the black hole, with the efficiency factor
$\epsilon _s $. In the standard picture
for the formation of a stellar mass black hole, a core of baryons is
first burned to heavy elements, and the subsequent collapse to a black
hole may release little more energy. In this case the
efficiency factor could be as small as $\epsilon _s\sim 0.009$, which
is the binding energy released as starlight. It could also be as large
as
$\epsilon _s\sim 0.03$ if the collapse proceeded through a
protoneutron star as an intermediate state. It cannot be much
larger, however,  without violating the constraints from the radiation
energy density (see \S 2.7) and the relic supernova neutrino flux
at Super-Kamiokande (Fukugita \& Kawasaki 2003).

One way to estimate the mass density in the massive black
holes in the nuclei of galaxies uses the correlation of the black hole
mass with the bulge luminosity. A convenient approximation to the
relation, for
B-band luminosities, is (Gebhardt et al. 2000; Ferrarese 2002;
see also Kormendy \& Richstone 1995)
\beq
M_\bullet /m_\odot = 10^{-2.0\pm 0.3} L_{\rm bulge}/L_\odot
.\label{eq:mbullet}
\eeq
FHP estimate that the fraction of the B-band luminosity density in
ellipticals and S0 galaxies is 0.24, and the fraction in the bulges of
spheroids is 0.14. The products of equation~(\ref{eq:mbullet}) with the
luminosity fractions and the luminosity density in
equation~(\ref{eq:luminositydensity})  gives the mass density
parameters in massive black holes,
\beqa
&&\Omega _\bullet (\hbox{early}) = 10^{-5.6\pm 0.3}, \nonumber\\
&&\Omega _\bullet (\hbox{late}) = 10^{-5.9\pm 0.3}.
\label{eq:omegabh}
\eeqa
Salucci et al. (1999) give a consistent, but slightly larger value.

For early-type galaxies we can use the tight
relation between the
black hole mass and the bulge or spheroid velocity dispersion
(Merritt \& Ferrarese 2001; Tremaine et al. 2002). The
Sheth et al. (2003) estimate of the velocity dispersion function for
early-type galaxies is
\beq
dN = \phi _\ast
\left(\sigma\over\sigma _\ast\right) ^\alpha
{\beta\over\Gamma (\alpha /\beta )} {d\sigma\over \sigma }
e^{-(\sigma /\sigma _\ast )^\beta },
\eeq
with $\alpha = 6.5$, $\beta = 1.93$, $\sigma _\ast = 89$ km~s$^{-1}$,
$\phi _\ast = 0.0020$ Mpc$^{-1}$. The Tremaine et al. (2002) estimate
of the
black hole mass-velocity dispersion relation is
\beq
M_\bullet = B(\sigma/\sigma _h)^a,
\eeq
with $B = 1.3\times 10^8m_\odot$, $a=4.0$, and $\sigma _h=200$
km~s$^{-1}$.
The product of the two expressions, integrated over $\sigma$, gives the
mean mass density,
\beq
\rho _\bullet = B\phi _\ast
{\Gamma ((\alpha + a)/\beta )\over\Gamma (\alpha /\beta )}
\left(\sigma _\ast\over\sigma _h\right) ^a.
\eeq
The numerical result,
\beq
\Omega _\bullet (\hbox{early}) = 10^{-5.9},\label{eq:omegabhalt}
\eeq
is close to but  smaller than the more direct estimate in
equation~(\ref{eq:omegabh}). Although the formal uncertainty in
equation~(\ref{eq:omegabhalt}) is smaller it rests on the condition
that the Sheth et al. galaxies are a fair sample of the early-type
galaxies, which will require careful debate.\footnote{The velocity
function of Sheth et al. gives $\langle\sigma^4\rangle ^{1/4}=180$
km s$^{-1}$, compared to our estimate of the characteristic velocity
dispersion, $\sigma_*=200-220$ km s$^{-1}$, in early-type galaxies.
Perhaps this is related to the difference.} Thus in the inventory we
quote equation~(\ref{eq:omegabh}).

\subsubsection{Quasar Luminosities and Remnants}

So\l tan (1982) and Chokshi \& Turner (1992) have considered the
relation between the rate of radiation of energy by quasars and AGNs and the accumulation of mass in the quasar  engines, which are assumed to be massive black holes in the centers of galaxies. In this repetition of the calculation we take the number of quasars per unit luminosity and comoving volume to be
\beq
L_\ast {dn\over dL} = {L_\ast\Phi (L_\ast )\over (L/L_\ast )^\alpha +
(L/L_\ast
)^\beta },
\eeq
where, from Croom et al. (2004), $\alpha = 3.31$, $\beta = 1.09$,
\beq
L_\ast\Phi (L_\ast ) = 1.81\times 10^{-6} \hbox{ Mpc}^{-3},
\eeq
and the present characteristic luminosity is
\beq
L_\ast = 6.7\times 10^{10}L_B(\odot) .
\eeq
In the Croom et al. luminosity evolution model this luminosity evolves as $ L_\ast(z)\propto 10^{1.39z -0.29z^2}$ to redshift $z=2.1$,
the deepest redshift used in the Croom et al. analysis. The peak of the observed comoving quasar number density is at $z\simeq 2.5$, and at $2.5<z<5$ the density varies about as $n\propto e^{-1.5z}$ (Fan et al. 2001, Fig. 3). As a convenient approximation to this behavior we adopt the Croom et al. (2004) luminosity evolution at $z<2.1$, constant comoving luminosity density from $z=2.1$ to $z=3$, and negligibly small luminosity at larger redshifts. 

The integral $\int  dL\,  L\, dn/dL$ is the comoving luminosity density, and the time integral multiplied by the bolometric correction is the net comoving density of energy released. 
Using the Elvis et al. (1994, Table 17) bolometric correction factor BC~$\equiv L_{\rm bol}/(\nu L_\nu)|_{\rm 4450\AA}=12$,  
and ignoring  the difference between B and $b_J$ passbands (since the quantity that concerns us, $\nu L_\nu$ for quasars, is close to flat), we estimate that the integrated energy density released by the quasars is
\beq
\Omega _{\rm QSO em}= 8\times 10^{-8}.
\label{eq:quasarradiation}
\eeq
This uses the solar luminosity,
\beq 
\nu L_\nu(\odot)=2.22\times 10^{33}\hbox{ erg s}^{-1}, 
\eeq
at $\lambda = 4450$ \AA .

Before comparing this estimate to the accumulated mass in black holes let us check consistency with the integrated background radiation. Equation~(\ref{eq:quasarradiation}) with the Elvis et al. bolometric corrections indicates that the integrated background from quasars at  $100\mu <\lambda <1000$~\AA\ is $\Omega\sim 3\times 10^{-8}$, or 1\% of the total (eqs. [\ref{eq:optical}]  plus [\ref{eq:fir}] below). To estimate the expected X-ray background we imagine the radiation from the quasars all is emitted at effective redshift $z=2$. In this simple model the present energy density per logarithmic interval of frequency at 2~keV is
\beq
\nu\Omega _\nu (2\hbox{ keV}) = 
{\Omega _{\rm QSO em}\over 1+z}
{\nu _e L_{\nu _e}\over\int _0^\infty L_\nu d\nu},
\label{eq:est}
\eeq
where on the right-hand side $h\nu _e = 2(1+z)$~keV.
The mean spectrum in figure~10 in Elvis et al. (1994) for radio-quiet quasars, with the bolometric factor, indicates that $\nu _eL_{\nu _e}= \int _0^\infty L_\nu d\nu /50$ at rest frame energy $h\nu _e=6$~keV. These numbers give the present energy density per logarithmic interval of photon energy $\nu\Omega _\nu =10^{-9.3}$ at 2~keV. The measured value of the X-ray background is (De Luca \& Molendi 2004) $\nu\Omega _\nu = 10^{-8.8}$ at 2 keV, about three times what is indicated by equations~(\ref{eq:quasarradiation}) and~(\ref{eq:est}). Since about 80\% of the X-ray background at 2~keV is resolved (Mushotzky et al. 2000; Worsley et al. 2004), these numbers allow room for a significant population of optically faint quasars. 

We turn  now to the efficiency $\epsilon _n$ for production of electromagnetic radiation in the accumulation of the present mass in quasar remnants. If $\epsilon _n$ is small the estimate of the integrated mass added to the black holes by the observed energy production by quasars and AGNs is 
$\Delta\Omega _{\rm BH} =\Omega _{\rm QSO em}/\epsilon _n$
 (eq.~[\ref{eq:quasarradiation}]).
The ratio of this expression to the mass density in massive black holes (the sum of entries 5.6 and 5.7) is our estimate of the radiation
efficiency,
\beq
\epsilon _n= 0.02.
\label{eq:qsoefficiency}
\eeq
This is one fifth of the commonly discussed value, $\epsilon _n\sim 0.1$. 
Since our estimate of the X-ray background from optically identified quasars is one third of the measured value, it may be that  equation~(\ref{eq:quasarradiation}) is low by a factor of about three and equation~(\ref{eq:qsoefficiency}) accordingly low by a like factor. A closer check of consistency of the idea that the massive black holes in the centers of galaxies are the quasar remnants awaits advances in surveys of optically faint quasars in broader ranges of wavelength and redshift and better understanding of the quasar emission mechanism. We may also hope that future work will establish the natures of the sources of the harder X-ray background and the possible relevance to the black hole mass budget.  

\begin{deluxetable}{llccr}
\tabletypesize{\scriptsize}
\tablecaption{Heavy Element Masses}
\tablewidth{0pt}
\tablehead{\colhead{} &  \colhead{Objects}  & \colhead{Mean
Metallicity}  & \colhead{Composition} & \colhead{$10^5\Omega _Z$} }
\startdata
1 & main seq. stars & $\langle Z\rangle$ & solar & $3.2$ \\
2 & substellar objects & $\langle Z\rangle$ &solar & $0.2$ \\
3 & white dwarfs& 1 & C+O & $36$ \\
4 & cool gas & $\langle Z\rangle$ & solar & $1.3$ \\
5 & clusters & $Z_\odot /3$ & solar & $1.1$ \\
6 & warm plasma & $Z_\odot /30$ & solar &  $2.5$\\
7 & neutron stars\tablenotemark{a} & 1 & (Fe) & $5$ \\
8 & stellar mass black holes\tablenotemark{a} & 1 & (Fe) & $6.8$ \\
9 & massive black holes\tablenotemark{a} & $Z_\odot$ & (solar) & $0.005$
\enddata
\tablenotetext{a}{vanished}
\end{deluxetable}

\subsection{Nuclear Binding Energy}

\subsubsection{Heavy Element Abundances}

We consider here the binding energy released by nuclear burning in
stars. We normalize the heavy element abundances to the Solar mass
fractions
in hydrogen, helium, and heavy elements,
\beq
X_\odot = 0.71,\quad Y_\odot = 0.27,\quad Z_\odot = 0.019.
\label{eq:solarabundances}
\eeq
The ratio $Z/X=0.027$ is derived (BP2000)
as the initial solar value from
$(Z/X)_{\rm solar~surface}=0.0230$ of Grevesse \& Sauval (2000),
and $X$ is also the initial value of BP2000.
The metallicity in star populations is
correlated with
the galaxy luminosity. An average over the Schechter luminosity
function of the
Kobulnicky \& Zaritsky (1999) correlation of the oxygen abundance with
the B-band galaxy luminosity, taking $12+ \log({\rm O/H})_\odot = 8.83$
(Grevesse \& Sauval 2000)\footnote{Recent work on the solar heavy
element
abundance suggests a significantly lower oxygen abundance,
      $12+ \log({\rm O/H})_\odot = 8.69$ (Allende Prieto, Lambert \&
Asplund 2001),
but if the heavy element abundances of other elements are scaled down
in a similar manner,
as indicated by the same team, our net result is not affected. In fact,
Bahcall and Pinsonneault (2004) give $(Z/X)_{\rm solar~surface}=0.0176$
based on the new abundance, which leaves our result unchanged.}
as the zero point, indicates that the mean metallicity in galaxies is
\beq
\langle Z\rangle =0.83(1\pm0.3) Z_\odot=0.016\pm0.003.
\label{eq:meanmetallicity}
\eeq

Table 3 lists our estimates of the density parameters belonging to the
mass
in heavy elements in several categories of objects. The entry for main
sequence stars in the first line is the product of $\langle Z\rangle$
with the sum of the density parameters in entries 3.3 and 3.4
in the inventory, and the second line uses entry 3.8. Since the main
elements in white dwarfs are carbon and oxygen, with thin hydrogen
and/or
helium layers that typically amount to $\la 0.1\%$ of the mass,
we enter
in the third line the density parameter from entry~3.5. We assign the
cool gas in entries 3.9 and 3.10 the same mean metallicity  as the
stars (eq.~[\ref{eq:meanmetallicity}]). The metallicity of the
intracluster plasma (entry 3.2) is observed to
be about one third of solar (Mushotzky \& Loewenstein 1997;
Fukazawa et al. 1998; White 2000), as indicated in Table~3. We suppose
the
intergalactic plasma (entry 3.1) may have metallicty about 3\%\ of
Solar. This likely is larger than the metallicity in plasma in the
voids (Penton, Stocke \&\ Shull 2004), and smaller than the metallicity
in the plasma observed as absorption line systems around galaxies
(Sembach et al. 2003; Churchill, Vogt, \& Charlton 2003), and perhaps
is a reasonable factor-of-three compromise. We assume that neutron
stars and stellar mass black hole form by the collapse of an
iron core, and that massive black holes grew by the accretion of
matter with about the Solar heavy element abundance. These heavy
elements are entered  in Table~3, but they are now sequestered
   from the inventory.

Our estimate of the total production of heavy elements, including those
that have been lost in neutron stars and black holes,  is
\beq
\Omega _Z=5.7\pm1.2\times 10^{-4}.
\eeq
The heavy elements in white dwarfs amount to about 65\%\ of the total.
The matter in this large reservoir is liberated only on the rare
occasions of Type Ia supernovae.

The model for the rate of  Type Ia supernovae
is uncertain; we consider the widely accepted Whelan \&\ Iben (1973)
binary white dwarf picture and use a  simple model for the supernova
rate,
\beqa
R_{\rm SN~Ia}(t) &=&A\int^t_{t_f}dt'\psi(t')
\int^{8m_\odot}_{m_{\rm min}} dm {dN\over dm}\times\cr
& &\exp{(-[t-t'-\Delta t]/\tau)},
\eeqa
where $A$ is the normalization determined by the empirical
Type Ia supernova rate at zero redshift, which is obtained
  from the three surveys mentioned earlier (eq.~[\ref{eq:obssnerate}]),
\beq
R_{\rm SNIa}(t_0)=0.0027^{+0.0017}_{-0.0008}~
(100{\rm yr})^{-1}\hbox{\rm Mpc}^{-3},\label{eq:obssnIrate}
\eeq
and
$\Delta t(m)=13(m/m_\odot)^{-2.5}~{\rm Gyr}+\delta$
is the time for the formation of white dwarfs plus the time delay to
form the Roche-lobe contact, and we take $\delta=0$ to 1~Gyr.
The minimum mass is chosen to be
$m_{\rm min}={\rm max}[3m_\odot,\{(t-t')/13{\rm Gyr}\}^{-0.4}]$,
where 3$m_\odot$ corresponds to the white dwarf mass 0.7$m_\odot$
according to eq. (\ref{eq:mwd-mms}) so that the Chandrasekhar limit
is met.
This is the model taken by Madau et al. (1998) (see also
Gal-Yam \& Maoz 2004). We adopt $\tau=4$ Gyr to account for a
moderate increase (by a factor of $2\pm1$) of the observed SNIa
occurence to $z=0.6$ (Pain et al. 2002).\footnote{A short time scale
is in conflict with the Type Ia supernova rates in early type galaxies.
   With $\tau=4$~Gyr, the SN fraction in early type galaxies is 0.30 at
    $z\approx 0$, which is consistent with the observed value,
$0.35^{+0.15}_{-0.10}$.   Here we have identified stars formed at
$z>0.7$ as an early population that is partitioned into early-type
galaxies and bulges of disk   galaxies according to the FHP fractions
of bulge luminosities. The
    fraction drops to $<0.25$ if $\tau=3$ Gyr.
   The observed rates of Type Ia
supernovae in morphologically separated galaxies seem to be
proportional to the $r$ band luminosity
density; the luminosity density in E/S0 galaxies is 31\%
of the total (Nakamura et al. 2003). This would suggest a close to
constant supernova rate.}
The effective time span for the cumulative occurrence of Type Ia
supernovae
normalised to the rate at $z=0$ is
\beqa
T_{\rm eff}&=&R_{\rm SNIa}(t_0)^{-1}\int^{t_0}_{t_f}R_{\rm SNIa}(t) dt
\cr
&=& 15\pm3\hbox{ Gyr},
\eeqa
for our model of star formation
(eqs.~[\ref{eq:sfrhistory}] and~[\ref{eq:sfrearly}] with
$z_f=2.5$ to 4). The cumulative comoving number density of supernova
in this model is $4.0\times 10^5$~Mpc$^{-3}$.
Since we estimate that the
number density of white dwarfs created is $8\times 10^7$~Mpc$^{-3}$,
this would mean that about 1\% of the white dwarfs have been disrupted.
If
$0.7m_\odot$ of
$^{56}$Ni is produced in each supernova (e.g., Branch 1992), the mass
density of iron-group elements produced by type Ia supernovae is
$\Omega_{\rm Fe}=2.0\times 10^{-6}$.
This is supplemented with
$\Omega_{\rm Fe}=3.6\times 10^{-6}$
   from Type II+Ib/c supernovae (see eq.~[\ref{eq:netsneproduction}]),
which produces
0.075$m_\odot$ of iron per event
(Arnett 1996; Weaver, Zimmerman \& Woosley 1978).
These estimates indicate that the total density parameter in the iron
group elements is
\beq
\Omega_{\rm Fe}=6\times 10^{-6},\label{eq:fe}
\eeq
of which about 60\% is from Type II
supernovae.

This estimate can be compared to the product of the mass in heavy
elements not  locked up in stellar remnants (the sum of lines 1, 2, and
4 to 6 in Table~3),
\beq
\Omega_Z'=0.8\pm0.25 \times 10^{-4}\ .
\label{eq:activez}
\eeq
with the iron group mass fraction (Grevesse \& Sauval 2000),
\beq
\Omega_{\rm Fe}=0.077\Omega_Z'=6.3\times 10^{-6}.
\eeq
The two approaches give a consistent picture for the origin of iron and
the supernova rates.

The growth of the abundance of heavy elements is accompanied
by the accumulation of helium, apart from the heavy elements that enter
white dwarfs, neutron stars and stellar mass black holes. The estimate
of $\Delta Y/\Delta Z$  in equation (\ref{eq:dydz}), applied to all
entries in Table~3 except 3, 7 and 8, is our estimate of the present
helium mass fraction in excess of primeval,
\beq\Delta\Omega _Y = {\Delta Y\over \Delta Z}\Omega_Z'
=1.7\pm1.0\times 10^{-4}.\label{eq:deltaomegay}\eeq
This is consistent with our estimate from the products of stellar
evolution (eq.~[\ref{eq:checkonhelium}]).

There are in the literature many analyses of the stellar production of 
heavy elements (see for example Calura \& Matteucci 2004 and references 
therein). Our results are generally consistent  for relevant entries. 
Important differences are our inclusion of heavy elements in white 
dwarfs, which dominate the entry, and our consideration of iron in the 
progenitors of supernovae. These components are crucial to the 
consideration of the balance between the nuclear binding energy stored 
in the elements and the energy released in radiation.

\subsubsection{Nuclear Binding Energies}

Our definition of the nuclear binding energies of the heavy elements
differs from the usual practice in  tables of nuclei. Because we are
interested in the release of energy in the formation of the heavy
elements, we calculate the binding energy with respect to free protons
and electrons.

We write the energy released in the production of the heavy elements
present in the interstellar medium, and in stars when they formed, as
\beq
\Omega _{\rm NB,\, Z}= (0.0081 + 0.0071\,\Delta Y/\Delta Z )\Omega _Z,
\label{eq:omeganb}
\eeq
where 0.0081 is the energy generation efficiency factor for the
solar composition.
The nuclear binding energy in substellar objects and in
diffuse matter, in entries 6.2, 6.4, and 6.5 in Table~1, is computed
   from this
equation with the heavy element masses in Table~3. For main sequence
stars,
we add to equation~(\ref{eq:omeganb}) the nuclear binding energy
associated with the helium that has been produced in
the stars, which we take to be on average 5\% of the star mass, that
is, half the helium a star produces while it is on the main sequence.
The sum is our estimate of the nuclear binding energy in stars in entry
6.1.

Entry 6.3, for white dwarfs, is
\beqa
\Omega _{\rm NB,\, wd} &=& [0.0080(1 - Y_p) +
0.0010Y_p]\Omega_Z\nonumber\\
&=&10^{-5.6}.
\eeqa
The second term in parentheses accounts for the primordial helium
abundance (eq.~[\ref{eq:primevalhelium}]). The value of $\Omega_Z$ in
the first line is taken from line 3 of Table~3. This component amounts
to about $40$\% of the nuclear binding energy. The large amount of
nuclear burning in the stellar  giant and supergiant phases is
discussed in connection with Table 5 below.

The nuclear binding energy in the matter out of which a neutron star
formed was converted to gravitational binding energy by the
dissociation of the heavy elements during a supernova. The nuclear
binding energy released in the formation of the heavy elements that
were part of the raw material for a  neutron star added to the
radiation background, of course, but that accounting now belongs in
category~ 5. The total entered in the inventory for category~6
accordingly is about 15\% smaller than the sum of the nuclear
binding energies in all entries in
Table~3. The nuclear binding energy to compare to the energy required
to produce the radiation background is the full sum, including
iron core progenitors for neutron stars and black holes,
\beq
\Omega_{\rm NB}=-5.7\pm 1.3\times 10^{-6}.\label{eq:starnucsyne}
\eeq
We discuss the relation to the energy density in radiation in the next
subsection.

\begin{deluxetable}{cc}
\tabletypesize{\scriptsize}
\tablecaption{Radiation Background}
\tablewidth{0pt}
\tablehead{ \colhead{\qquad$\lambda$ ($\mu$)\qquad}
& \colhead{\qquad$\nu I_\nu$ (nW m$^{-2}$ sr$^{-1}$)\qquad}  }
\startdata
0.10 & 0.60\tablenotemark{a}\\
0.30 & $13\pm 8$\tablenotemark{b} \\
0.37 & $24\pm 8$\tablenotemark{c} \\
0.55 & $18\pm 8$\tablenotemark{b} \\
0.81 & $24\pm 9$\tablenotemark{b} \\\
2.2 & $23\pm 9$\tablenotemark{d} \\
3.5 & $12\pm 3$\tablenotemark{d} \\
\enddata
\tablenotetext{a}{Henry (1999)\quad $^{\rm b}$Bernstein et al. (2002)}
\tablenotetext{c}{Matilla (1990)\quad $^{\rm d}$Wright \& Reese (2000)}
\end{deluxetable}

\subsection{The Radiation Backgrounds}

The cosmic energy density in electromagnetic radiation is thought to be dominated by mildly redshifted starlight, at $\lambda\sim 1\mu$, and a far infrared peak at $\lambda\sim 100\mu$ that is produced by the absorption  and reradiation of starlight and the light from AGNs
(Hauser \& Dwek 2001 and references therein). The energy densities at radio and X-ray to $\gamma$-ray wavelengths are much smaller, but they are useful measures of high energy processes, as is the gravitational wave background. Neutrino production might be counted as part of the radiation backgrounds, but we find it convenient to enter neutrinos in a separate category.

\subsubsection{The $\lambda\sim 1\mu$ Background}

Observations of the optical to near infrared  extragalactic background
light
that report positive detections are summarized in Table 4. They
suggest that the surface brightness per logarithmic interval
in frequency is about constant at $\nu I_\nu = 20\pm 5$
nW~m$^{-2}$~sr$^{-1}$ in the range 3500~\AA \ to 3.5~$\mu$.
This corresponds to energy density
\beq
\Omega _{\rm opt}=2.3\pm 0.6\times 10^{-6},\label{eq:omegaoptical}
\eeq
in the optical to near infrared.  In view of the technical difficulty of
these observations, equation~(\ref{eq:omegaoptical}) may conservatively
be taken as an upper limit. Integrated galaxy number counts give
surface brightness typically one third of the entries in Table~4 (Madau
\& Pozzetti 2000; see also Hauser \& Dwek 2001, Table 3),
\beq
\Omega _{\rm opt}=0.9\pm 0.2\times 10^{-6}.\label{eq:optgalcount}
\eeq
This might
be considered a lower limit.

We have a check from the energy density computed as a time
integral of the luminosity density,
\beq
u = \int {d\lambda\over\lambda } \int _0^{t_o} {dt\over 1 + z}\lambda
(t) {\cal L}_{\lambda (t)}.
\label{eq:integraloflumden}
\eeq
The integral is from high redshift to the present world time, $t_o$.
The integrand is the luminosity per logarithmic interval of wavelength
and comoving volume, corrected by the redshift factor, $1+z=1/a(t)$,
and evaluated at the redshifted wavelength $\lambda (t) = a(t)\lambda
_o$, where the observed wavelength is $\lambda _o$ and the expansion
parameter  $a(t)$ is normalized to unity at the present epoch. In the
flat cosmological model the integral is
\beqa
&&u = H_o^{-1}\lambda _o{\cal  L}_{\lambda _o}I,\nonumber\\
&&I = \int _0^1{da\over\sqrt {\Omega _m/a^3 + 1 - \Omega _\lambda}}
\int {d\lambda {\cal L}(\lambda ,z)\over \lambda _o{\cal L}_{\lambda
_o}}.\qquad
\label{eq:integrate}
       \eeqa
       We take the shape of the present cosmic spectrum $\lambda {\cal
L}_\lambda$ as a function of wavelength from Figure 13 in Blanton et
al. (2003), and we normalize to the luminosity density at $\lambda\sim
1\mu$ in equation~(\ref{eq:luminositydensity}). We approximate the
evolution of the comoving luminosity density by extrapolating the
Rudnick et al. (2003) power law fits to the evolution in the rest frame
U, V and B bands, in the form
       \beq
       {\cal L}(\lambda ,z) \propto (1+z)^\beta, \
       \beta = 0.93(4400\hbox{\AA}/\lambda )^{2.5}.
       \eeq
We truncate the integral at $z=3$, the limit of the Rudnick et al.
measurements. In this model the dimensionless integral in
equation~(\ref{eq:integrate}) is $I=1.35$. (If the comoving luminosity
density were constant the integral would be $I=0.82$.) The result is
\beq
\Omega _{\rm opt}= 1.5\times 10^{-6}.\label{eq:opticalintegral}
\eeq

For the inventory we adopt
\beq
\Omega _{\rm opt} = 1.6\pm 0.7\times 10^{-6}.\label{eq:optical}
\eeq
The error spans the estimates based on measurements of the surface
brightness of the sky (eqs.~[\ref{eq:omegaoptical}]), the source counts
(eq.~[\ref{eq:optgalcount}]) and the luminosity density
(eq.~[\ref{eq:opticalintegral}]).

\subsubsection{The Far Infrared Background}

The COBE DIRBE (Hauser et al. 1998) and FIRAS (Fixsen et al. 1998)
experiments detect the extragalactic radiation background at
$\lambda\ga 100\mu$. The integral for $\lambda>125\mu$ is
$u=14$ nW~m$^{-2}$~sr$^{-1}$ (Fixsen et al. 1998).
Extending the integration to $100\mu$ might reasonably be expected to
add
2 nW~m$^{-2}$~sr$^{-1}$ to this value. The density parameter for the
sum is
\beq
\Omega_{\rm FIR}=0.8\pm 0.2\times 10^{-6}.\label{eq:fir}
\eeq
This is entry 7.2 in the inventory. It seems to be a believable lower
bound on the energy density in the far infrared. The radiation
measurements allow room for a comparable amount of  energy at
$30\la\lambda\la 100~\mu$
(Finkbeiner, Davis \& Schlegel 2000; see also Hauser \& Dwek 2001).
However, this amount of energy shortward of 100$\mu$ wavelength would
distort the TeV $\gamma$-ray spectrum from the extragalactic source
Mrk501 (Quinn 1996;  Aharonian et al. 1999), by the absorption due to
$e^+e^-$ pair production (Kneiske, Mannheim \& Hartmann 2002;
Konopelko et al. 2003). Thus it appears that equation~(\ref{eq:fir}) is
close to the total in the far infrared.

The optical and infrared light from quasars is about one percent of the starlight and far infrared backgrounds (\S 2.5.4).

\subsubsection{A Check: Nuclear Burning}

We may compare the estimate of the present energy density in the
optical through the far infrared against what would be expected from
the energy stored in the heavy elements
(eq.~[\ref{eq:starnucsyne}]),  and what would be expected from the
picture for stellar evolution. We comment on the former here and the
latter in \S 2.8.1.

The sum of equations (\ref{eq:optical}) and (\ref{eq:fir})
(entries 7.2 and 7.3), corrected for  the redshift energy loss
factor in equation (\ref{eq:redshiftloss}), is an estimate of the
nuclear energy required to produce the observed radiation,
\beq
\Omega_{\rm photon}= 4.8 \pm 1.4 \times 10^{-6}.
\label{eq:betoradiaton}
\eeq
This number may be compared to the nuclear binding energy released
by nucleosynthesis. The estimate in
equation~(\ref{eq:starnucsyne}), reduced by 7\% to take account of the
energy carried away by neutrinos, is
\beq
\Omega_{\rm photon,\, nuc.}=5.3 \pm 1.1\times
10^{-6}.\label{eq:nuccheck}
\eeq
The difference, 10\%, is well within our uncertainties.
It might be relevant to note that the difference between
equations~(\ref{eq:betoradiaton}) and~(\ref{eq:nuccheck}) would be
increased if we adopted the estimate of the optical background from
source counts.

\subsubsection{The X-ray --- $\gamma$-Ray Background}

Entry 7.4 in Table~1 is the integral of the radiation background
spectrum compiled by Gruber et al. (1999) over
the energy range 3~keV to 100~GeV. The largest contribution to the
integral is at photon energy $\sim 30$~keV, but the convergence at high energy is slow because the energy per logarithmic interval of
frequency varies about as $\nu\Omega _\nu\sim \nu^{-0.1}$. The measurement at 3~keV by De Luca \& Molendi (2004) is about 1.3 times the Gruber et al. value. We have not attempted to adjust the entry for this more recent result because we do not know its significance for the spectrum at higher energies. 

\subsubsection{The Radio Background}

Longair (1995), following Bridle's (1967) study, concludes that  the
brightness temperature of the isotropic radio background at 178~MHz,
after correction for the thermal component, is $T=27\pm 7$~K. This
corresponds to
\beq
u_\nu=6600(1\pm 0.25)\left(\frac{\nu}{\rm 1GHz}\right)^{-0.8} {\rm
Jy~sr}^{-1},
\label{eq:radio}
\eeq
assuming the canonical power law spectrum (see also Peacock 1995).
Longair also argues that this radiation is dominated by galaxies.

Equation~(\ref{eq:radio}) may be compared to  the sum of the radio
source counts in the 9CR radio survey
at 15 GHz, in the observed range of flux densities, 0.01 to 1 Jy
(Waldram et al. 2003). This sum gives
$u_\nu=5900$ Jy sr$^{-1}$ when scaled to 1 GHz with the power index of
0.8.
The radio source counts at 8.4~GHz by Fomalont et al. (2002),
summed over 10 $\mu$Jy to 1 Jy, give  $u_\nu=4700$ Jy sr$^{-1}$
at 1 GHz. A comparable result is obtained  from the count at
40~GHZ (as summarized in Figure~13 of Bennett et al. 2003b).
Haarsma \& Partridge (1998) argue that the integral over the
counts likely converges at  $S\sim 1\,\mu$Jy. We conclude that
the measurements are reasonably concordant with
equation~(\ref{eq:radio}),
within $\sim$0.2 dex.

The integral of equation~(\ref{eq:radio}) slowly diverges at
short wavelengths. We adopt, as an operational definition, a cutoff at
$\lambda=1$mm ($\nu =300$~GHz), and we count the contribution at
shorter wavelengths as part of the FIR background in our inventory.
There is a natural cutoff at long wavelength, at $\nu\sim$3~MHz (Simon
1978). Integrating eq. (\ref{eq:radio}) over this wavelength range, we
obtain
\beq
\Omega_{\rm radio}=5\times 10^{-11}.
\label{eq:radioenergy}
\eeq

To understand the relation to other energy entries,
we may attempt an alternative estimate by directly
summing the contributions of known radio galaxies. The luminosity
function of radio galaxies at zero redshift is now
reasonably well known by virtue of the correlation of a large
NRAO VLA Sky Survey at 1.4~GHz (NVSS; Condon et al. 1998)
with galaxies in optical catalogues (Condon, Cotton \&
Broderick 2002 [UGC vs. NVSS]; Machalski \& Godlowski 2000
[LCRS vs. NVSS]; Sadler et al. 2002 [2dF vs. NVSS]).
The luminosity function is written as the sum of two components,
weak radio emitters that represent normal galaxies with
star forming activity, and strong emitters that mostly consist of
subsets of giant elliptical galaxies and AGNs.
The former activity is ascribed to electron acceleration
in Type II (and Ib/c) supernova remnants.
At close to zero redshift the integral over the luminosity function
gives
luminosity densities
\beqa
&&{\cal L}_g = 1.5\times 10^{19}\hbox{W Hz}^{-1}\hbox{ Mpc}^{-3},
\nonumber \\
&&{\cal L}_a=3.7\times 10^{19}\hbox{ W Hz}^{-1}\hbox{ Mpc}^{-3},\qquad
\label{eq:radiolumdensity}
\eeqa
at 1.4 GHz (Condon et al. 2002). The second component, AGNs, is
dominated by the most luminous sources, and so is subject to sampling
fluctuations. Haarsma et al. (2000)
show that the evolution of the first component in
equation~(\ref{eq:radiolumdensity}) is fast to
$z\sim 1$. The evolution they derive is consistent with the model
for the evolution of the star formation rate in
equation~(\ref{eq:sfrhistory}). Condon (1992) gives
the relation between the radio emissivity and the star formation rate.
After adjustment for the overall constraint from the star formation
rate density derived from the local H$\alpha$ luminosity density, as
discussed in
\S 2.3.2, and our reference IMF, the relation reads
\beqa
&&u_\nu=2.5\times 10^{21}\left(\frac{\nu}{\rm GHz}\right)^{-0.8}
\nonumber\\
&& \qquad +
2.6\times 10^{20}
\left(\frac{\nu}{\rm GHz}\right)^{-0.1}~{\rm W~Hz}^{-1},\qquad\
\label{eq:radiosed}
\eeqa
per 1 $m_\odot$ yr$^{-1}$ of star formation.
The first component represents synchrotron radiation,
primarily from supernovae, and
the second represents bremsstrahlung from HII regions.
The bremsstrahlung contribution becomes more important above 25 GHz.
The integral over the star formation history to $z_f=3$
(with the redshift energy loss)
and the frequency range $\nu = 3$~MHz to 300~GHz
gives $\Omega_{\rm radio}=3.3\times 10^{-11}$, where
the contributions from synchrotron radiation and bremsstrahlung
are in the proportion 0.4:0.6 in the frequency range that concerns us.

The evolution of the strong radio sources appears not to be very fast.
The luminosity functions of
Sadler et al. (2002) and Machalski \& Godlowski (2000) up to
$z\sim 0.2$ suggest a slow evolution compared with that of
the weak emitter component, at least at low redshifts.
We take the evolution factor derived from
high redshift strong radio galaxies by Willott et al. (2001),
$(1+z)^{2.6}E(z)$ with $E(z)=[\Omega_m(1+z)^3+\Omega_\lambda]^{1/2}$,
and we assume the canonical synchrotron spectrum.  The integral over
time and frequency yields $\Omega_{\rm radio}=2.0\times 10^{-11}$.
The sum of the two components is
\beq
\Omega_{\rm radio}=5\times 10^{-11}.
\label{eq:radioenergy2}
\eeq
The consistency with equation~(\ref{eq:radioenergy}) suggests
that this simple model for the radio sources is a useful approximation.

\subsubsection{Gravitational Radiation}

We consider two sources of gravitational radiation, the merging of stellar mass binaries and the formation of massive black holes in the centers of galaxies. 
The rate of production of gravitational radiation by
stellar mass binaries in the Milky Way depends on the
very uncertain distribution of properties of close binary
systems that contain white dwarfs, neutron stars, and
black holes. The  largest contribution to the mean
gravitational radiation luminosity of the Milky Way is
thought to be from the merging of binary neutron stars, as
indicated in Fig. 12 in Schneider et al. (2001). 

Shibata \& Ury{\= u} (2002) and Faber et al. (2002) find that in the merging of a binary neutron star system the energy emitted as gravitational radiation is 
\beq
\epsilon _{\rm GWnn}\simeq 0.005
\label{eq:nnefficiency}
\eeq
times the mass of the binary. A major uncertainty is the behavior of the system after contact. Gravitational radiation emitted during the transient formation of an oscillating neutron star might double the number in equation~(\ref{eq:nnefficiency}). 

If we adopt the estimate of the rate of merging of binary neutron
stars in the Milky Way, ${\cal R} \approx 80$~Myr$^{-1}$,
that is based on the three known systems that will merge within
the Hubble time (Kalogera {\it et al.} 2004a, b),
then we find that the product with the radiation energy released (eq.~[\ref{eq:nnefficiency}]), the
effective number density of Milky Way galaxies
(eq.~[\ref{eq:MWgaldensity}]), and 85~Gyr/2
(eqs.~[\ref{eq:redshiftloss}] and [\ref{eq:timespan}]) is the
gravitational radiation energy density $\Omega _{\rm NS,gw}= 10^{-9}$.
The merger rate is higly uncertain: the spread of central values
in the models in Table~1 in Kalogera {\it et al.} (2004b) is $10^{+0.7}_{-1}$.
On taking account of the upward uncertainty in equation~(\ref{eq:nnefficiency}),
we are led to allocate an error of a full order of magnitude up or down
in entry 7.5.

The merging rates for neutron star-black hole and black hole-black hole binaries are even more uncertain. The indication from the analysis of  Schneider et al. (2001) is that mergers involving black holes add about 10\%\ to the gravitational radiation luminosity. 

In the low frequency regime ($\sim$ mHz) the luminosity is dominated by compact white dwarf binaries, as shown by 
Hils, Bender \& Webbink (1990). The integration of the 
gravitational lumiosity given in their Table 7 with the number of 
close white dwarf binaries in the Milky Way normalised to 
$3\times 10^6$ (after a suggested reduction by a  
factor of ten from the simple theoretical estimate of 
the abundance of these objects) gives the luminosity $3\times 10^{39}$ erg~s$^{-1}$. The product with the effective number
density of Milky Way galaxies and the effective time span with redshift correction gives $\Omega _{\rm WD, gw}\sim 10^{-10}$. This is at the lower end of the suggested range of uncertainty in the value of $\Omega _{\rm NS,gw}$.

Our estimate of the gravitational radiation released by the formation of the massive black holes in the centers of galaxies is based on the analysis by Baker et al. (2001), which indicates that the gravitational radiation energy released by the merging of equal mass black holes is $\epsilon _{\rm ngw}\simeq 0.03$ times the final black hole mass $M$, and that the frequency of the radiation is on the order of $0.1M^{-1}$, or about $10^{-4}M_9$~Hz when the mass is measured in units of 
$10^9m_\odot$. We take the redshift at peak production to be $z_n\sim 2.5$, about the peak of quasar activity. Then the present energy density of the gravitational radiation from the formation of the massive black holes, with mass density $\Omega _n=10^{-5.4}$, is
$\Omega _{\rm ngw}\sim \epsilon _{\rm ngw}\Omega _n/(1+z_n)\sim
10^{-7.5}$. This may be compared to the bound on the gravitational wave
energy density from  pulsar timing measurements, $\Omega_g< 10^{-8}$ at
frequency $\sim 10^{-8}$~Hz (Kaspi, Taylor \& Ryba 1994; Thorsett \&
Dewey 1996; Lommen \& Backer 2001). The bound is about 30\%\ of our 
estimate of $\Omega _{\rm ngw}$, but there is no inconsistency because 
the expected peak frequency is considerably larger than the frequency 
where the pulsar constraint is  tight. The spectrum of the
gravitational radiation from merging massive black holes, and the
puzzle of how merging black holes get close enough that energy loss by
gravitational radiation drives merging within a Hubble time, are
discussed by Hughes et al. (2001),
Jaffe \&\ Backer (2003) and Wyithe \& Loeb (2003).

An entry in the inventory for possible sources of low frequency
gravitational radiation in the early universe, as from inflation or
phase transitions, seems premature.

\subsection{Products of Stellar Evolution}

It is in principle straightforward to compute the integrated outputs of
stellar evolution --- energy, neutrinos, helium, and heavy elements ---
given models for the IMF, the star formation history, and stellar
evolution. Since the details of the results of stellar evolution
computations are not easily assembled, we use approximate estimates
by procedures similar to those developed in \S 2.3.1 and 2.3.2 for the
stellar population and its evolution. The
results add to the checks of consistency of our estimates of the
stellar production of helium and heavy elements and the resulting total
energy release, and are used to estimate the inventory entries for the
neutrino cosmic energy density.

\begin{deluxetable}{lcccc}
\tabletypesize{\scriptsize}
\tablecaption{Stellar energy production\tablenotemark{a}}
\tablewidth{0pt}
\tablehead{ \colhead{stage of stellar evolution}  &
\colhead{$0.08-1m_\odot$}  &
\colhead{$1-8m_\odot$} & \colhead{$8-100m_\odot$} &
\colhead{sum}}
\startdata
main sequence & 0.11  & 0.20  & 0.12  & 0.43\\
H shell burning &     & 0.18  & 0.29  & 0.48\\
core evolution &      & 0.05  & 0.04  & 0.09\\
\noalign{\vspace{3pt}}\tableline\noalign{\vspace{3pt}}
sum           & 0.11  & 0.43  & 0.46  & 1.00\\
\enddata
\tablenotetext{a}{Normalised to $\Omega=5.3\times 10^{-6}$.}
\end{deluxetable}

\subsubsection{Stellar Evolution}

Most stars with masses $m<1\, m_\odot$ are still on the main sequence.
We assume that on average 5\% of the hydrogen in these subsolar stars
has been  consumed, with energy production efficiency 0.0071. Most of
the stars with masses $m>1\, m_\odot$ have already undergone full
evolution and left compact remnants, while the fraction
$\sim (m/m_\odot)^{-2.5}$ is still on the main sequence. We
assume that in the latter stars on average 5\% of the hydrogen has been
consumed, as for subsolar stars. For the evolved stars we do not
attempt to follow the details of nuclear burning and mass
loss. Instead, we adopt estimates of the nuclear fuel consumed or mass
lost in a few discrete stages of evolution, in a similar fashion to the
approach used in \S 2.3.1 to tally stellar remnants.

When the amount of hydrogen consumed in stellar burning is 10\% of the
mass of a star it leaves the main sequence. In the model in \S 2.3.1,
stars with masses in the range $1<m<8 m_\odot$ eventually produce white
dwarfs that mainly consist of a carbon-oxygen core. In standard stellar
evolution models, hydrogen burning extends outward to a shell after
core hydrogen exhaustion, and helium burning similarly continues in a
shell after core helium exhaustion. That leaves a CO core with the mass
given
by equation (\ref{eq:mwd-mms}). The helium layer outside the CO core is
thin for $m<2.2m_\odot$, for which helium ignition takes place
as a flash, but in stars with masses $m>2.2m_\odot$ a significant
amount of
helium is produced outside the
core, transported by convection to the envelope, and liberated. A
$5\,m_\odot$ star liberates
$0.4\,m_\odot$ of helium and produces a core with mass $0.85\,m_\odot$
(e.g., Kippenhahn, Thomas \&\ Weigert 1965). We model the helium
production as a
function of the initial star mass $m$ as
\beq
m_{\rm He}=0.23(m-2.2)+0.63,
\label{eq:lowmasshelium}
\eeq
in solar mass units. The energy production
in hydrogen shell burning is the product of this mass with the
post-stellar hydrogen mass fraction, 0.71
(eq.~[\ref{eq:solarabundances}]), and the efficiency factor, 0.0071.
Helium burning in the core
produces energy with efficiency factor 0.0010.

For stars with masses $m>8 m_\odot$ we adopt the helium
yield from Table 14.6 of Arnett (1996),\footnote{We prefer
eq.[\ref{eq:core_he_mass}] which is derived from tabulated values,
rather than Arnett's parametrisation,
$m_{\rm He}=0.43m-2.46$.}
which we parametrise as
\beq
m_{\rm He}=0.69m-3.9.
\label{eq:core_he_mass}
\eeq
This connects to equation~(\ref{eq:lowmasshelium})
at $8.7\,m_\odot$.
We take the CO core mass as a function of the initial stellar mass from
Arnett (1996):
\beq
m_{\rm CO}=0.28m-2.20, \quad m>13\,m_\odot .
\label{eq:cocoremass}
\eeq
We use an interpolation of eqs.~[\ref{eq:mwd-mms}] and
[\ref{eq:cocoremass}] for stellar masses between 8 and
13$m_\odot$.
The energy release is $0.0071\times0.71$
per unit mass for He production, and 0.0014 for CO core formation
and the further heavy element production.

The energy output obtained by integration over the IMF and PDMF is
\beq
\Omega_{\rm stellar~E}=5.3\times 10^{-6}\ .
\label{eq:stellare}
\eeq
The partition into each phase of stellar evolution and stellar mass
range
is given in Table 5, where the numbers are normalised to equation
({\ref{eq:stellare}). About 60\% of the energy is produced in the
evolved
stages.

The estimate of the total energy generation in
equation~(\ref{eq:stellare}) is in satisfactory agreement with our
estimate of the nuclear binding energy,
$\Omega_{\rm BE}=5.7\times 10^{-6}$ (eq.~[\ref{eq:starnucsyne}]),
and the energy production required to produce our estimate of the
present radiation energy density, $\Omega_{\gamma + \nu }=(5.1\pm
1.5)\times 10^{-6}$ (eq.~[\ref{eq:betoradiaton}] corrected for neutrino
emission, as discussed in \S 2.8.2).
That is, our models for stellar formation and evolution are
consistent with our estimates of the accumulation of heavy elements and
the stellar background radiation.

In our models for stellar formation and evolution the mass of helium
that has been liberated to interstellar space is
\beq
\Delta\Omega_Y\approx 2.1\times 10^{-4}\ .\label{eq:checkonhelium}
\eeq
This may be compared to our estimate from the production of heavy
elements and the associated production of helium,
$\Delta\Omega_Y=1.7\pm 1\times 10^{-4}$ (eq.~[\ref{eq:deltaomegay}]).
The heavy element production is calculated in a
similar way. The current wisdom is that a neutron star or
black hole is left at the end of the evolution of a
star with mass $m>8\,m_\odot$, and the rest of the
mass is returned to the interstellar medium.
On subtracting the remnant mass (indicated in Table~2) from the total
heavy element mass produced, we find that in our model the heavy
element production is
\beq
\Omega_Z\approx 5\times 10^{-5}\ .
\label{eq:zprod}
\eeq
The more direct estimate in equation (\ref{eq:activez}) is
$\Omega_Z'\simeq 8\times 10^{-5}$.

We suspect the checks on helium
and heavy element production in equations~(\ref{eq:checkonhelium})
and~(\ref{eq:zprod}) are as successful as could be expected. In
particular, equation~(\ref{eq:cocoremass}) for the CO core mass is not
tightly constrained. One must consider also the possibility that some
supernovae do not produce compact remnants, as is suggested by the
cases of Cas A and SN1987A. The maximum heavy element production
when there are no remnant neutron stars or black holes is three times
the value in equation~(\ref{eq:zprod}), and larger than $\Omega_Z'$
(eq.~[\ref{eq:activez}]).

\subsubsection{Neutrinos from Stellar Evolution}
\label{sec:nufromhburn}

We need the fraction $f_\nu$ of the energy released as neutrinos by the
various processes of nuclear burning. In stars with masses $m<1.4\,
m_\odot$, energy generation in hydrogen burning is dominated by
the slow reaction
$p+p\rightarrow d+e^++\nu_e$. This produces neutrinos with mean energy
0.265 MeV, which amounts to $f_\nu =0.020$ times the energy generated
in helium synthesis. In a  Solar mass star electron capture of
$^7$Be adds 0.005 to the fraction $f_\nu$. The neutrino energy emission
fraction is larger in higher
mass ($m>1.4\,m_\odot$) main-sequence stars in which  the CNO cycle
dominates. In this process, the neutrinos produced in the $\beta^+$
decays of $^{13}$N and $^{15}$O carry away the energy fraction $f_\nu
=0.064$. The fraction increases to 0.075 for $m\ga 2m_\odot$, when the
subchain $^{15}$N-$^{16}$O-$^{17}$F-$^{17}$O-$^{14}$N starts dominating
and neutrinos are produced by the
beta decay of $^{17}$F. This sidechain also dominates during shell
burning.

The integral of these factors over the stellar IMF
normalised to the present-day mass density $\Omega_{\rm
star}=0.0027$ (eq.~[\ref{eq:omegastars}]), together with our
prescription for energy generation in Table~5, yields the neutrino
energy production,
\beq
\Omega_{\nu, {\rm ms} } =0.31\times10^{-6}.\label{eq:nums}
\eeq

The temperatures and densities that are reached up to carbon burning
are low enough that there is negligible neutrino energy loss from
neutrino pair production processes.

The temperatures after carbon burning are high enough that the
neutrino energy loss dominates, that is, $f_\nu\simeq 1$
(Weaver et al. 1978). Thus we may
take it that the extra binding energy $\Delta$(BE)
of the elements heavier than
$^{20}$Ne with respect to the binding energy of carbon represents the
neutrino energy emitted in the late stages
of stellar evolution. On multiplying the heavy element mass
(the sum of entries 1, 2 and 4 to 9 in Table 3) by $\sum
(Z_i/Z)\Delta({\rm BE})_i$ (where $\langle\Delta({\rm
BE})\rangle=0.0004$ for the solar mix of element abundances, and $\sum
Z_i/Z=0.35$ is from
Grevesse \& Sauval (2000), we obtain
\beq
\Omega_{\nu ,{\rm pairs}}=0.03\times 10^{-6}.\label{eq:nupairs}
\eeq

The sum of equations~(\ref{eq:nums}) and~(\ref{eq:nupairs}) is 7\%\ of
the total energy production (eq.~[\ref{eq:stellare}]). The present
energy density of the stellar neutrinos, in entry 8.1 in Table~1, is
the product of this sum with the redshift loss factor $\sim 0.5$
(eq.~[\ref{eq:redshiftloss}]).

\subsubsection{White Dwarfs and Neutron Stars}

Most of the gravitational energy liberated in white dwarf formation
goes to
neutrinos, and in the latest stage to X-rays. Since the latter is small
the contribution to the neutrino energy density (entry 8.2 in the
inventory) is the product of the gravitational binding energy in entry
5.3 with the redshift factor.

More than 99\% of the energy released in
core collapse also is carried away by neutrinos. Thus we similarly
obtain entry 8.3 by multiplying entry 5.4 by the redshift factor.

\subsection{Cosmic Rays and Magnetic Fields}

This estimate is based on the rate of production of cosmic rays in the
Milky Way galaxy. We start from the radio structure. In their analysis
of the Haslam et al. (1982) measurements, Beuermann, Kanbach, \&
Berkhuijsen (1985) write the radio luminosity distribution at 408~MHz
as the sum of thin and thick components. We model each distribution, in
an approximation to their results, as
\beq
{\cal E}(\vec r)={\cal E}(0) e^{-(|z|/z_{\rm eq} + r/l}).
\label{eq:radiostructure}
\eeq
In the thick component the scale heights in the disk and perpendicular
to the disk are
\beq
l = 3.1\hbox{ kpc}, \qquad z_{\rm eq} = 1.4\hbox{ kpc},
\eeq
scaled to Solar galactocentric distance $8$~kpc (see Appendix).
The total luminosity
in the thin component in this model may be neglected, but the thin
component is a significant contribution to the local luminosity
density, ${\cal E}_\odot$: the local thick component fraction is ${\cal
E}_\odot ({\rm thick})/{\cal E}_\odot = 0.55$. We assume the energy
density in cosmic rays is proportional to the luminosity density at
408~MHz. A more detailed model would take account of the spatial
variation of the magnetic field strength, but that will be left for
future studies. We normalize to the local energy density, 1.8 eV
cm$^{-3}$, in relativistic cosmic rays (Webber 1998). In these
approximations the energy of cosmic rays in the galaxy is $10^{56.1}$
erg. We take the
cosmic ray mean life in the galaxy to be $2\times 10^7$~yr
(Garcia-Munoz, Mason \& Simpson 1977; Yanasak et al. 2001). The ratio
is a measure of the cosmic ray luminosity, $10^{41.3}$ erg s$^{-1}$.
If the acceleration of cosmic rays is mainly due to shocks of Type II
and Ibc supernovae,\footnote{No type Ia supernova remnants are
known to give strong radio sources (Weiler et al. 1986).
This implies that Type Ia supernovae contribute little to the
acceleration
of cosmic rays.} following the conventional wisdom
(Ginzburg \& Syrovatskii 1964),
the cosmic ray luminosity is proportional to the star formation
rate. Thus the product of this luminosity with the effective time
(eq.~\ref{eq:timespan}) and the  redshift factor
(eq.~[\ref{eq:redshiftloss}]) is an estimate of the contribution of the
Milky Way to the present energy in cosmic rays. The product with the
effective number density of Milky Way galaxies
(eq.~[\ref{eq:MWgaldensity}]) is our estimate of the present energy
density in cosmic rays from normal galaxies,
\beq
\Omega_{\rm CR}=10^{-8.3}.\label{eq:crdensity}
\label{eq:cr}
\eeq

The radio luminosity of the Milky Way offers a check of this
calculation. The Beuermann et al. (1985) luminosity of the Milky Way,
$5.5\times 10^{21}$ W~Hz$^{-1}$ at $\nu =408$~GHz, scaled to 1.4~GHz by
the $\nu ^{-0.8}$ power law (neglecting the small thermal
bremsstrahlung component), and multiplied by the effective number
density of Milky Way galaxies (eq.~[\ref{eq:MWgaldensity}]), is ${\cal
L}_g=9\times 10^{18}$ W~Hz$^{-1}$~Mpc$^{-3}$, about half the Condon et
al. (2002) measurement of the mean luminosity density of the galaxies
(eq.~[\ref{eq:radiolumdensity}]). That is, the evidence is that the
Milky Way gives a  reasonably good measure of the synchrotron
luminosity density of the galaxies, and hence of the cosmic ray
luminosities of galaxies. This is not a very direct check of the
assumed universality of the cosmic ray lifetime in the source galaxy,
of course.

A similar energy may be present in the magnetic field.
If the leakage of cosmic
rays approximated a fluid flow, magnetic field would
leak  into intergalactic space with the cosmic rays. Application of
equipartition to equation (\ref{eq:cr}) would suggest that the
cosmic rms magnetic field strength is
\beq
      B_{\rm IGM} \sim 3\times 10^{-8}\hbox{ Gau\ss}.\label{eq:bigm}
\eeq
If the estimate of the local ratio of the
cosmic ray  to magnetic field energy density,  8:1
(the magnetic field corresponding to 3~$\mu$Gau\ss),
applied to the intergalactic
medium, it would lower equation~(\ref{eq:bigm}) by a factor of three.

The product of the integrated production of supernovae with the
characteristic kinetic energy, $E_{\rm KE}\simeq 1.6\times 10^{51}$
erg, released in a supernova
(Arnett 1996), is an estimate of the integrated kinetic energy
production per comoving volume,
\beq
\Omega_{\rm SN~KE}\simeq10^{-7.3}.
\label{eq:snke}
\eeq
If the fraction $\epsilon _{\rm cr}$ of this energy were placed in
relativistic intergalactic particles, the present energy density in
this component, taking account of the redshift factor, would be
\beq
\Omega_{\rm SN~cr}\simeq10^{-7.6}\epsilon _{\rm cr}.
\label{eq:sncr}
\eeq
This is consistent with the picture that a fraction $\epsilon _{\rm
cr}\sim 0.2$ of the kinetic energy liberated by supernovae has been
deposited in cosmic rays (eq.~[\ref{eq:cr}]), which eventually
become intergalactic.

Cosmic rays might gain energy from shocks produced by
streaming motions of magnetized
warm plasma in the vicinity of galaxies (Loeb \& Waxman 2000).
However, since the energy available in the plasma
(eq.~[\ref{eq:warmgaske}] below) is less than the energy liberated by
supernovae (eq.~[\ref{eq:crdensity}]), this process is not likely to
substantially affect the energy in cosmic rays.

What is the AGN contribution to the cosmic ray energy density? Though
AGNs are not major sources at optical wavelengths, their contribution
to the radio background is comparable to that of the larger number of
normal galaxies (eq.~[\ref{eq:radiolumdensity}). This is an indication
of the importance of AGNs for high energy processes. If the cosmic ray
energy production by AGNs were proportional to
the production of synchrotron radio emission, the energy of cosmic rays
  from AGN's (including radio elliptical galaxies) would be comparable
to, or even larger than, that from the normal galaxies, and our
estimate in
eq. (\ref{eq:crdensity}) would have to be doubled. It is important also
that AGNs may accelerate cosmic rays to higher energies than do normal
galaxies.

We use equation~(\ref{eq:crdensity}) in the inventory.  The lower error
flag for the entry in Table~1 is based on the evidence that the Milky
Way gives a fair measure of galaxy radio luminosities. The larger upper
error flag reflects the possibly significant roles of the intergalactic
magnetic field and  AGNs.

\subsection{Kinetic Energy in the IGM}

Intergalactic baryons have peculiar velocities that are driven by the
gravitational field of the dark matter distribution and by
nongravitational interactions. The former is part
of the primeval energy in category 4 in the inventory. The latter is
expected to be most important near the virialized regions of galaxies,
where the
streaming motion has been largely transformed by shocks into thermal
energy at  the temperature $T\sim 2\times 10^6$~K associated with the
nominal velocity
dispersion $\sigma =160$~km~s$^{-1}$ around $L\sim L_\ast$ galaxies
(eq.~[\ref{eq:virial}]). This shocked matter may be responsible for the
     O~VI absorption lines in the  Local Group (Sembach et al. 2003; Cen
et al. 2001). The product of the internal energy
per unit mass of plasma at this temperature with the mass fraction in
entry 3.1a is
\beq
\Omega_{\rm KE}\simeq 10^{-8.0}.
\label{eq:warmgaske}
\eeq

If an appreciable fraction of the kinetic energy produced by supernovae
(eq.~[\ref{eq:snke}]) were deposited as kinetic energy in galactic
haloes this energy would be dissipated by hydrodynamic processes rather
than the cosmological redshift, but still may make a significant
addition to equation~(\ref{eq:warmgaske}), heating the IGM relative
to dark matter.

Baryonic matter well outside the nominal virialized regions of galaxies
and larger systems --- at distances greater than about $r_{200}$ from
clusters of galaxies and  $r_v\sim 200$~kpc from $L\sim L_\ast$
galaxies --- is observed as Ly$\alpha$ absorption systems (Penton,
Stocke \&\ Shull 2004). The primeval peculiar motions of this matter
are perturbed by photoionisation that produces kinetic temperatures on
the order of $10^4$ K.  The product of this kinetic energy per unit
mass with the intergalactic mass fraction (entry 3.1b)
is about one percent of the kinetic energy in the warm intergalactic
component (eq.~[\ref{eq:warmgaske}]).

We  use equation~(\ref{eq:warmgaske}) for category 10. We caution,
however, that a comparable amount of kinetic energy may be deposited by
supernova winds.

\subsection{Electrostatic Energy}

The effect of the electrostatic interaction on the binding energies of
atomic nuclei relative to free protons and electrons is taken into
account in category 6, and the electrostatic contribution to the
binding energies of white dwarfs is (in principle) part of category 5.
The molecular binding energy in objects ranging from dust to asteroids
that are held together by the electrostatic interaction deserves
separate mention.

The molecular binding energy relative to free atoms in condensed matter
is roughly 1 eV per atom. The product of this mass fraction, $\sim
10^{-10}$, with entry 3.12 is
\beq
\Omega _{\rm BE, molecular}\sim -10^{-16}.\label{eq:asteroidbinding}
\eeq
This is the binding energy of condensed matter physics outside strongly
self-gravitating systems. We refrain from entering it in Table~1
because the estimate is so small and uncertain, but we offer for
comparison the binding energy of the electrons in atoms.

The electrostatic binding energy of the electrons in an O VI atom is
1.6~keV, and the addition of the other five electrons in a neutral
oxygen atom increases the binding energy by only 27 percent. That is,
it is a reasonable approximation to ignore the states of ionization of
the heavy elements and the electrostatic binding energy of neutral
hydrogen and helium. The sum over the heavy element masses in entries
1, 2, 4, 5, and 6 in Table 3, weighted by the neutral  atomic binding
energies of 15 cosmically conspicuous elements, is
\beq
\Omega _{\rm BE, atoms} = -10^{-9.7}.
\eeq
This is some six orders of magnitude larger than the molecular binding
energy in equation~(\ref{eq:asteroidbinding}), and comparable in value
to the smallest entries in Table~1.

\section{Discussion}

The convergence of the cosmological tests from redundant constraints on
the global cosmological parameters offers a good case for the
$\Lambda$CDM Friedmann-Lema\^\i tre cosmology as a useful approximation
to the real world. This is a necessary condition that our compilation
of the energy
inventory is a meaningful exercise.

We adopt the working assumption that the galaxies are useful tracers of
mass, in the sense that inventory entries derived under this assumption
are good approximations to reality. This assumption has been widely
questioned. However, the recent evidence from weak lensing
(eqs.~[\ref{eq:delatsigma}] to~[\ref{eq:weaklensing}]), with the older
evidence from the galaxy relative velocity dispersion at separations
$\sim 100$~kpc to 1~Mpc, and the consistent picture for the virialized
parts of field galaxies (eq.~[\ref{eq:virial}]), offer what seems to be
a reliable case for the use of galaxies as mass tracers. The approach
certainly is not exact, and working out more accurate measures of the
mass distribution on the scale of galaxies remains an important and
fascinating challenge.

Our other conventions for the cosmology are less controversial but
may have to be adjusted. For simplicity, we have adopted a fixed
distance scale (eq.~[\ref{eq:hnot}]). We have not presented the scaling
of inventory entries with the distance scale, which can be somewhat
complicated; this issue is best revisited when the uncertainty in the
distance scale is better understood. Our adopted value and formal
uncertainty for the matter density parameter, $\Omega _m$
(eq.~[\ref{eq:omegam}]), is in the generally accepted range,
but there is reason to suspect that it is somewhat overestimated
(\S 2.1). Again, this is a issue to revisit when the constraints have
improved. Our value for the baryon mass density agrees with what is
derived from the CMBR anisotropy and the measurements of the primeval
deuterium abundance, but it is larger than what is indicated by
the primeval helium abundance. We have suggested in \S
2.2.2 that the problem may be with the technical difficulties of the
helium measurements, but we will be following the advances in this
subject with interest. It would not be
surprising to see the discovery of richer physics in the dark sector,
and with it an increase in the variety of entries in category 1, and
possibly also some adjustment of other parts of the inventory that
depend on the dark sector physics. One should also bear in mind that
the alternative pictures of structure formation that were under
discussion a decade ago, such as cosmic strings, could be operating as
subdominant perturbations to the $\Lambda$CDM model. Further progress
in testing the model for structure formation will inform our ideas on
whether the entries that depend on the theory of structure formation
are likely to require adjustment. Each of these issues may be pointing
to  revisions to the inventory outside our stated errors. On the
other hand, the successful network of tests of the cosmology and the
model for structure formation leads us to expect that the general
framework presented in Table~1 is not likely to change.

We have argued that $6\pm 1\%$ of the baryons are in stars and stellar
remnants (eqs.~[\ref{eq:baryonomega}] and~[\ref{eq:omegastars}] with
$h=0.7$). This small fraction  is analyzed in considerable detail in the
inventory, in entries that are supported by a network of
tests. The estimate in \S 2.7.1 of the mass density in stars
uses the optical luminosity density and the stellar-mass-to-light
ratio. The luminosity density is checked by measurements of the optical
to near infrared intergalactic radiation energy density, and by the
measured
galaxy counts (\S 2.7.1). The comparison is not tight, but it
shows that the radiation energy density likely is known to $\pm 0.3$
dex. The conversion from the luminosity density to the stellar mass
density depends on the model for the stellar initial mass function. The
model we use is checked by the reasonable agreement with the ratio of
the white dwarf to subsolar main sequence star densities
at the low mass end, and with the global Type II supernova rate
at the high mass end (2.3.1, 2.3.2). The model for the star formation
history is based on the time history of the H$\alpha$ luminosity
density, which is broadly consistent with other measures of the
evolution of the star formation rate density (subject to the
considerable uncertainty in extinction in the UV). The model is checked 
by
consistency with the accumulated mass in stars (after correction for
mass loss;  eq.~[\ref{eq:halphasfr}]). Yet another check involves the
accumulated mass density
in heavy elements. The release of nuclear binding energy in the heavy
elements, corrected for the loss of radiation energy by the
cosmological redshift (eq.~[\ref{eq:redshiftloss}]), is in satisfactory
agreement with the present radiation energy density at near optical and
far infrared wavelengths (\S 2.7.3). This test would fail if there were
a substantial amount of radiation energy at wavelengths
$1\mu<\lambda< 100\mu$; improved measurements will be of
considerable interest. This test also depends on the stellar production
of helium, which is discussed in \S 2.2.2, in connection with the
constraint on the baryon mass density from the primeval helium
abundance, and in \S 2.8.1, from an analysis of the products of stellar
evolution. The results seem reasonably consistent. The network of
checks is complicated, but that is what lends credence to the results.

It is well to pause to consider Arp's (1965) cautionary remark, that
there may be extragalactic objects with sizes too small to be readily
distinguished from stars --- an example is the quasi-stellar objects
Arp mentions in a note added in proof --- or with surface brightnesses
too low to be readily seen against the foreground --- an example is the
intracluster light in the Virgo cluster (Arp \& Bertola 1971). The
quasar remnants could have contained a significant baryon mass, if the
mass conversion efficiency $\epsilon _n$ had been close to unity
(eq.~[\ref{eq:bhbindingenergy}]), but we now know that that is
difficult to reconcile with the integrated quasar energy emission
(eq.~[\ref{eq:quasarradiation}]). There are low surface brightness
objects (e.g McGaugh, Schombert, \& Bothun 1995), but the surface
brightness of the extragalactic sky shows that they cannot largely
affect our estimate of the mean luminosity density (\S 2.7.1).
Substellar objects might be counted as part of Arp's cautionary remark,
but they are detectable  in nearby galaxies by weak gravitational
lensing (or MACHOs, as discussed by Alcock et al. 2000 and references
therein). That is, a half century ago it was not clear that it
is feasible to establish a fair observational census of the stars. Now
it appears that the observational conditions allow it: the closure of
our inventory suggests that missing or unknown components cannot be
energetically very significant.

The baryon number density --- excluding baryons that may have been
sequestered prior to light element production  --- seems to be reliably
constrained, but the states of most of the baryons are not yet
observationally well documented.
The picture that galaxies trace mass leads us to expect that about half
the dark matter is gathered near and within the virialized parts of the
galaxies (eq.~[\ref{eq:galaxydarkmass}]). Entries 3.1 and 3.2 in
Table~1 are based on the assumption that the baryons are similarly
placed, in the
diffuse states observed as hot plasma in clusters of galaxies and in
warm and hot plasma in and around groups of galaxies
(eq.~[\ref{eq:igmfraction}]). This has not yet been convincingly
observationally demonstrated. We have not made much use of the
predicted states of the baryons from numerical simulations, because we
do not know how to judge the reliability of the predictions on
relatively small scales. As the computations and their observational
checks improve this will become clearer, and the results may be
expected to improve this part of the inventory.

The advances in observations of the massive compact objects in the
centers of galaxies allow an improved test of the idea that these
objects are the black hole remnants of quasars and AGNs that are
powered by gravitational accretion. Our estimates of the mass density in black holes and the luminosity density in quasars require that the quasar mass conversion efficiency is $\epsilon _n\sim 0.02$ (eq.~[\ref{eq:qsoefficiency}]), close to the efficiency of production of gravitational radiation (\S 2.7.6). If the estimate of the quasar luminosity density has missed a significant population of optically faint AGNs then our value of $\epsilon _n$ is biased low. The quasar luminosity density is quite uncertain, as is the actual value of $\epsilon _n$, but it is encouraging that the present result is as close as it is to conventional ideas about the quasar emission mechanism. 

The efficiency of conversion of mass into electromagnetic radiation in
the formation of stellar mass black holes is constrained to be less
than a few percent  by the observed energy density in
electromagnetic radiation and neutrinos. The
interpretation of this constraint awaits development of the theory of
the astrophysical processes of formation of stellar mass black holes.
Our estimate of the gravitational binding energy released by core
collapse supernovae (entry 8.3) is comparable to the energy released in
the formation of the massive black hole quasar remnants, and it is not
very much less than the upper bound from neutrino detectors. Progress
here will be followed with interest.

The largest energy densities in the inventory are in dark matter and 
dark energy. Any empirical hints to the natures of these components 
will be followed with great interest. One possibility is that the 
observations of microlensing toward the Magellanic Clouds have detected massive compact dark objects (Alcock et 
al. 2000; Afonso et al. 2003). We have adopted the conservative  interpretation, that the 
lenses are baryonic --- stars and stellar remnants --- with standard 
mass functions, and
  that the high lensing rate is an accident of the
  distributions of source and lensing stars. A demonstration of the apparently simpler
  interpretation, that the lenses contribute some 20\%\ of the mass of 
the halo of the Milky Way, and are baryonic, perhaps white dwarfs, 
would considerably upset our considerations of the stellar baryon 
budget; a demonstration that the lenses are nonbaryonic would 
profoundly affect thinking about dark matter and the early universe.

  Though there are large gaps in our understanding of the cosmic energy inventory, most notably in the dark sector, there also is an 
impressively broad observational basis for many parts of the inventory. 
Ten years ago the basis was considerably smaller, and we can be sure 
that ten years from now the observations will allow more entries and a 
richer network of crosschecks of the interpretations. Progress will be 
uneven, of course, as one sees by considering the prospects for further 
checks
of the neutrino energy inventory. With the recent demonstration that
the solar neutrino luminosity agrees with standard physics augmented by
neutrino masses and mixing angles (Bahcall 2003 and references herein)
we may now be confident that the theory of neutrino production by
stellar nuclear burning is reliably tested even in the absence of a
direct observation of stellar neutrinos from remote stars.
The neutrinos produced by
plasma processes in white dwarf formation are, while an enormous
energy density, too soft for detection by any conceivable method: 
estimates of this component of the inventory will have
to continue to rely on the theory. Neutrinos released in stellar core
collapse have been detected (Hirata et al. 1987; Bionta et al. 1987),
and the integrated background of relic neutrinos from this process is 
expected to be within experimental reach. The
detection would complete another valuable check of the cosmic energy
transactions. 

Parts of our inventory can be improved on the basis of what is available within existing computations, or could be readily developed from existing work, but that we could not readily assemble. In particular,
the theoretical analyses of stellar remnants and the other products of
stellar evolution (\S\S 2.3.1 and 2.8), with special attention to the
white dwarfs that store so much of the heavy elements, will
have to be done better. Another example is the heavy element abundances, which can be classified by element type, as
oxygen  --- which has a specific spectroscopic importance --- and the
$\alpha$-elements and the iron group, and by environment, as white
dwarfs, main sequence stars, and interstellar and intergalactic space.

It is also worth stating explicitly that we may expect that, as in the past,  there will be qualitatively new experimental or observational developments --- perhaps the detection of gravitational radiation, or the identification of dark matter particles --- that substantially affect the directions of research and the 
development of the cosmic energy inventory.

\section{Concluding Remarks}

A half century ago we had a largely conjectural picture of the
large-scale structure of the physical universe, of its material
contents, and of the main process that drive transformations among the
states of matter and radiation. Now our world picture has a substantial
basis in experiments and observations and the attendant theories. This
has been an evolutionary process: none the entries in the inventory in
Table~1 requires a substantial departure from ideas that are under
discussion, and in many cases have been in the literature for decades.
The important new development is that we now have observational support
for the many entries in Table~1, and a network of tests that
demonstrate that a considerable part of the inventory is a believable
approximation.  Continued advances in the observational and theoretical
basis for the inventory surely will yield unexpected revisions and
additions; we are attempting to draw large conclusions from limited
observations of an exceedingly complex universe. However, the big
surprise at the moment is that it is now possible to find an inventory
with observational support for the largest $\sim 40$ forms of energy.

\acknowledgments

We have benefited by advice from Charles Alcock, Dave Arnett, John
Bahcall, Michael Blanton, Pierre Bergeron, Bruce Draine, Doug
Finkbeiner, Peter Goldreich, Andy Gould, Vicky Kalogera, Kim Griest, Craig Hogan, David Hogg, Guinevere Kauffmann, Julian Krolik, Jim Liebert, Geoff Marcy, Bruce Partridge, Martin Rees, Neill Reid, Aldo Serenelli, Masaru Shibata, Mike Shull, Gary Steigman, Paul Steinhardt, Stuart Shapiro, Tomonori Totani, Scott Tremaine, Simon White, and the referee.

PJEP thanks the Japan Society for Promotion of Science
for supporting his visit to Japan, where this work was initiated.
MF received support for this work from the Monell Foundation at the
Princeton Institute for Advanced Study, and a Grant in Aid from the
Japanese Ministry of Education at the University of Tokyo.

\appendix

\section{A Few  Astrophysical Quantities for the Milky Way}

The current best value for the distance to the Galactic center is
\beq
R_0=7.94\pm 0.42 \hbox{ kpc}, 
\label{eq:R0}
\eeq
 from measurements of the orbit --- proper
motion and redshift --- of a star close to Sgr A* (Eisenhauer et al. 2003). The product of this distance with the measurement of the proper motion of Sgr A$^\ast$ by Reid \& Brunthaler (2004) is 241~km~s$^{-1}$. This velocity corrected for the solar proper motion is our adopted value of the circular speed of the Milky Way at the Solar circle,
\beq
\Theta_0=234 \pm 13\hbox{ km s}^{-1}.
\label{eq:Th0}
\eeq
Traditionally, one invokes the Oort model to infer the circular speed, but Olling \& Dehnen (2003) find that the current best result depends substantially on the stellar populations used in the analysis. 

The circular velocity yields an estimate of the optical luminosity of 
the Milky Way via the Tully-Fisher relation.
To convert the optically measured 
maximum rotation velocity  $V_{\rm rot}$ to the 20\%
line width of HI measurements $W_{20}$, we use the relation
\beq
W_{20}/2=(0.93\pm0.02) V_{\rm rot}+(27\pm1.5) {\rm km~s}^{-1}.
\label{W20}
\eeq
derived from the data given by Mathewson \& Ford (1996). 
  From the Galactic rotation curve from HI and CO terminal velocities
(using Fig. 9.17 of Binney \& Merrifield 1998), we estimate that 
the maximum rotation velocity of the Milky Way is
$V_{\rm max}=1.03\Theta_0=241\pm13$ km s$^{-1}$. 
Identifying  $V_{\rm max}=V_{\rm rot}$,
we find 
\beq
W_{20}({\rm MW}) =502\pm30\hbox{ km s}^{-1}. 
\eeq

  From the COBE-DIRBE integrated light for the Milky Way in the
near infrared (J, K, L) wavelength bands, Malhotra et al. (1996) showed 
that the Milky Way is on the standard Tully-Fisher relation
in these wave length bands, slightly on the brighter
side but within the dispersion around the relation.
Assuming that the Milky Way has the standard color of spiral
galaxies, we may expect that the luminosity in the B band also 
satisfies the Tully-Fisher relation for the B band. Using the Sakai et 
al. (2000)
local calibration, 
\beq
M_{B_T^c}=-19.80-7.97(\log W_{20}-2.5),
\eeq
we find
\beq
L_{B^c}({\rm MW})=10^{10.74\pm 0.19}L_\odot ,
\label{eq:MWlum1}
\eeq
where we use $B_\odot =5.46$, and the error is dominated by 
the intrinsic dispersion
(0.43 mag) of the B-band Tully-Fisher relation. The superscript
$c$ stands for extinction-free quantities.

An alternative estimate of the B-band luminosity uses the
colour transformation applied to the DIRBE integrated K-band 
luminosity of the Milky Way. A comparison of the samples of 
Malhotra et al.(1996) and Sakai et al. (2000) indicates  $B_T^c-K^c=2.84\pm0.10$
for typical Sbc galaxies. Thus $K^c({\rm MW})=-23.95\pm0.25$ yields
$B_{\rm_T}^c=-21.1\pm0.3$, or
\beq
L_{B^c}({\rm MW})=10^{10.63\pm 0.13}L_\odot .
\label{eq:MWlum2}
\eeq
Combining the two, our recommended value is 
\beq
L_{B^c}({\rm MW})=4.6\pm1.4 \times 10^{10}L_\odot .
\label{eq:MWlum3}
\eeq
A value often
quoted in the literature,
$L_B^c=2.3\pm 0.6 \times 10^{10}L_\odot$ (van den Bergh 1988), is 1.5 
standard deviations below equation~(\ref{eq:MWlum3}).

The luminosity obtained here is the extinction-free value. When it
is compared with $L^*$ that appears in the luminosity function, we
must take account of internal extinction. For consistency we use the 
average
value of the internal extinction correction in the Sakai et al. (2000) 
local calibration,
$A_B^{\rm int}= 0.44$ mag, or an attenuation factor 1.5. This means that
$L_B(\rm MW)=1.40 L_B^*$ at $h=0.7$, and that the effective number density 
of Milky Way galaxies is the value in equation~(\ref{eq:MWgaldensity}), 
${\cal L_B}/L_B(\rm MW)=0.013 h^3 \hbox{ Mpc}^{-3}$.

\end{document}